\documentclass[a4article]{JHEP3}
\usepackage{amsmath}
\title{
Pohlmeyer reduction revisited\\[7pt]}
\author{J.~Luis Miramontes\\
Departamento de F\'\i sica de Part\'\i culas, and\\ Instituto Gallego de
F\'\i sica de Altas Energ\'\i as (IGFAE),\\Universidad
de Santiago de Compostela\\ 15782 Santiago de Compostela, Spain\\

E-mail: \email{jluis.miramontes@usc.es}}

\abstract{
A systematic group theoretical formulation of the Pohlmeyer reduction is presented. It provides a map between the equations of motion of sigma models with target-space a symmetric space ${\cal M}=F/G$ and a class of integrable multi-component generalizations of the sine-Gordon equation. When ${\cal M}$ is of definite signature their solutions describe classical bosonic string configurations on the curved space-time ${\fl R}_t\times {\cal M}$. In contrast, if ${\cal M}$ is of indefinite signature the solutions to those equations can describe bosonic string configurations on ${\fl R}_t\times {\cal M}$, ${\cal M}\times S^1_\vartheta$ or simply ${\cal M}$. The conditions required to enable the Lagrangian formulation of the resulting equations in terms of gauged WZW actions with a potential term are clarified, and it is shown that the corresponding Lagrangian action is not unique in general. The Pohlmeyer reductions of sigma models on ${\fl C}P^n$ and $AdS_n$ are discussed as particular examples of symmetric spaces of definite and indefinite signature, respectively.
}

\keywords{Sigma Models, Integrable Field Theories, Bosonic Strings}
\def\bfm#1{\boldsymbol{#1}}
\newcommand{\fl}[1]{\mathbb{#1}}    
\newcommand{\go}[1]{\mathfrak{#1}}    

\newcommand\blank[1]{#1}
\renewcommand\blank[1]{}

\begin{document}
\section{Introduction}
\label{Introduction}

Pohlmeyer reduction provides a map between the equations of motion of two-dimensional sigma models and a class of  
multi-component integrable generalizations of the sine-Gordon equation. 
It relies on the classical conformal invariance of sigma models that can be exploited to choose coordinates such that the components of the stress-energy tensor are constant; namely,
\begin{equation}
T_{++}=T_{--}=\mu^2
\label{StressTensor}
\end{equation}
together with $ T_{+-}=0$. The simplest examples, originally discussed by Pohlmeyer, are provided by the reduction of the $S^2=SO(3)/SO(2)$ and the $S^3=SO(4)/SO(3)$ sigma models, which yield the well-known sine-Gordon and complex sine-Gordon equations, respectively~\cite{Pohl} (see also~\cite{Pohl2}). This procedure has been generalised to the sigma models associated to generic symmetric spaces following a group theoretical approach that leads to the so-called symmetric space sine-Gordon (SSSG) equations~\cite{ECPn,EHCPn,SSSG2,SSSG3,SSSG4}~\footnote{In~\cite{HSG} the name `symmetric space sine-Gordon' was used to denote only a particular subset of SSSG theories described by gWZW actions with a potential term associated to cosets of the form $G/U(1)^p$, which were singled out by the condition of having a mass gap without exhibiting non-abelian global symmetries.}, whose integrability properties are very well established. In contrast, their Lagrangian formulation was a long-standing problem until Bakas, Park and Shin proposed their identification with the equations of motion of specific gauged Wess-Zumino-Witten (gWZW) actions modified by suitable potentials~\cite{BPS} (see also~\cite{Parkold,HMP,FMSG,HSG}). 

If we denote by ${\cal M}=F/G$ the target-space of the sigma model, the conditions~(\ref{StressTensor}) can be identified with the Virasoro constraints of bosonic string theory on the curved space-time ${\fl R}_t\times {\cal M}$ using the orthonormal gauge $t=\mu\tau$~\cite{Tseytlin2003}. Then, Pohlmeyer reduction provides a classical relation between (integrable) generalised sine-Gordon equations and bosonic string theory on curved space-times of that type~\footnote{A different string theoretical interpretation of Pohlmeyer reduction follows from the work of Lund and Regge~\cite{Lund}}. 
This alternative formulation has proved to be very useful in the study of the classical spectrum of string theory on curved space-times and, in particular, in the investigation of the AdS/CFT correspondence. For instance, applied to the subspaces ${\fl R}_t\times S^2$ and ${\fl R}_t\times S^3$ of $AdS_5\times S^5$, Pohlmeyer reduction relates the `giant magnons' of~\cite{HM} and the `dyonic giant magnons' of~\cite{CDO} to the soliton solutions of the sine-Gordon and complex sine-Gordon equations, respectively. More examples can be found in~\cite{MORE}. In a similar way, it seems likely that Pohlmeyer reduction will also be useful in the study of the recently proposed
duality between superstrings on $AdS_4 \times {\fl C}P^3$ and
$N = 6$ super Chern-Simons theory~\cite{NEW}.

The `stringy' interpretation in terms of ${\fl R}_t\times {\cal M}$ requires that $\mu^2>0$  in~(\ref{StressTensor}), which is the only possibility if ${\cal M}$ is a manifold of definite (positive) signature. 
This includes compact manifolds like the $n$-spheres $S^n=SO(n+1)/SO(n)$ or  the complex projective spaces ${\fl C}P^n=SU(n+1)/U(n)$, and noncompact ones like the hyperbolic $n$-spaces $H^n=SO(1,n)/SO(n)$.
In contrast, if ${\cal M}$ is of indefinite signature Polhmeyer reduction can also be performed with $\mu^2<0$. Then, it 
provides a relationship between  two-dimensional integrable equations and bosonic string theory on ${\cal M}\times S^1_\vartheta$, where the conditions~(\ref{StressTensor}) arise as the Virasoro constraints in the gauge  $\vartheta=\mu\tau$, with $\vartheta$ being the $S^1$ angular coordinate~\cite{GTseytlin1}. Important examples of manifolds of indefinite signature are the de~Sitter $dS_n=SO(1,n)/SO(1,n-1)$ and anti-de~Sitter $AdS_n=SO(2,n-1)/SO(1,n-1)$ spaces.
Furthermore, for symmetric spaces of this type, it is also possible to perform Polhmeyer reduction with $\mu^2=0$. Obviously, this case is different to the others with $\mu^2\not=0$, since the constraints~(\ref{StressTensor}) do not break conformal invariance. 
Nevertheless, it has a natural interpretation in terms of
bosonic string theory on ${\cal M}$ and, in fact, 
Pohlmeyer reduction with $\mu^2=0$ has already been used to construct classical bosonic string configurations on de~Sitter and anti-de~Sitter spaces in~\cite{DeVega,Jevicki}.

From the point of view of the original sigma model degrees of freedom, Pohlmeyer reduction amounts to a non-local  transformation of variables that breaks conformal invariance (provided that $\mu^2\not=0$) while preserving integrability and two-dimensional Lorentz invariance. Taking this into account, 
Grigoriev and Tseytlin~\cite{GTseytlin1,GTseytlin2} and Mikhailov and Sch\"afer-Nakemi~\cite{MikhailovSS} have recently proposed a generalization of Pohlmeyer reduction to reduce the $PSU(2,2|4)/Sp(2,2) \times Sp(4)$ supercoset model and, in this way, to find a novel, manifestly two-dimensional Lorentz invariant formulation of the full $AdS_5\times S^5$ superstring theory, which is known to be classically integrable~\cite{BPR}, alternative to the usual formulation in the light-cone gauge~\cite{MT}. The starting point of their proposal is the reduction of the bosonic part of the supercoset, which consists of two decoupled $AdS_5$ and $S^5$ sigma models. Using classical conformal invariance, the corresponding Virasoro constraints can be written as
\begin{equation}
T_{\pm \pm}^{(AdS_5)}=-\mu^2 
\quad {\rm and}\quad
T_{\pm \pm}^{(S^5)}=+\mu^2,
\end{equation}
which leads to two decoupled Pohlmeyer-reduced $AdS_5$ and $S^5$ sigma models. Then, the proposed new Lagrangian formulation of the $AdS_5\times S^5$ superstring theory is found by generalizing the approach of~\cite{BPS}, and it is provided by a gWZW action with a potential term coupled also to a set of two-dimensional fermionic fields. 

The implementation of the proposal of~\cite{GTseytlin1,GTseytlin2,MikhailovSS} requires a rather precise understanding of the relationship between the degrees of freedom of the original sigma model and those of the gWZW action with a potential term that describes the reduced model. For the sigma models which are relevant to the proposal, it has been explicitly worked out in~\cite{GTseytlin1} (see also~\cite{GTseytlin2,MikhailovSS}). However, although the results of that paper apply to a larger class of reduced sigma models, they are not general enough to describe the relationship between the degrees of freedom in all the possible cases and, in particular, in the reduced models associated to symmetric spaces or rank larger than 1~\footnote{The rank of a symmetric space $F/G$ is the dimension of the maximal abelian subspaces in the orthogonal complement of the Lie algebra ${\go g}$ of $G$ in the Lie algebra ${\go f}$ of $F$. It satisfies the bounds
$\mathop{\rm rank}(F)-\mathop{\rm rank}(G)\leq \mathop{\rm rank}(F/G) \leq \mathop{\rm rank}(F)$ (see~(\ref{RankBound}) and Table~\ref{RankTable}).
$S^n$, ${\fl C}P^n$ and $AdS_n$ are examples of symmetric spaces whose rank is~1.}.

The purpose of this paper is to provide a systematic group theoretical formulation of Pohlmeyer reduction that makes explicit the relationship between the equations of motion of sigma models with target-space a generic (bosonic) symmetric space, the corresponding SSSG equations, and the equations of motion of the gWZW actions that provide their Lagrangian formulation. 
It is organized as follows.
In Section~\ref{NLSM}, we summarise the construction of the nonlinear sigma model with target-space a symmetric space $F/G$. Its equations of motion 
can be written in terms of the currents $J_\pm=P_{\go p}(f^{-1}\partial_\pm f)$, where $f$ is a field that takes values in $F$, $P_{\go p}$ is the orthogonal projector on the complement of the Lie algebra ${\go g}$ of $G$ in the Lie algebra ${\go f}$ of $F$ (see~(\ref{SSalgebra})), and $\partial_\pm=\partial_\tau\pm \partial_x$. They imply that $\partial_\pm \mathop{\rm Tr}(J_\pm^n)=0$, which provides an infinite set of local chiral conserved densities.
Then, Pohlmeyer reduction amounts to constraining all those conserved densities to be constant. Namely, 
\begin{equation}
\mathop{\rm Tr}(J_\pm^n)= \text{\rm constant} \quad \forall \;n\geq2,
\label{GenConstraint}
\end{equation}
which includes 
the constraints~(\ref{StressTensor}) that correspond to $n=2$.
An early motivation for this characterization of Pohlmeyer reduction procedure can be found in~\cite{SSSG3}.

In Section~\ref{SSSG}, we shall solve those constraints for symmetric spaces of definite (positive) signature using the so-called `polar coordinate decomposition'. In this case, all the local chiral densities $\mathop{\rm Tr}(J_\pm^n)$ can be written as polynomials in terms of only $2\mathop{\rm rank}(F/G)$ `primitive' densities~\cite{EvansMountain}. Then, if $\mathop{\rm rank}(F/G)=1$, the only primitive densities are $T_{++}$ and $T_{--}$ and, consequently, the reduction gives rise to only one set of SSSG equations which, up to a classical conformal transformation, is equivalent to the equations of motion of the original sigma model. In contrast, if $\mathop{\rm rank}(F/G)>1$ the reduction involves more primitive chiral densities than just $T_{++}$ and $T_{--}$ and, for different choices of their constant values, the reduction procedure gives rise to rather different sets of SSSG equations. Their solutions correspond to bosonic string configurations on ${\fl R}_t\times F/G$ subjected to additional constraints.
The relationship of the SSSG equations with the non-abelian affine Toda equations~\cite{NAATls,NAATtau,NAATkir} is clarified in Section~\ref{NAAT}, and their Lagrangian formulation is discussed in Section~\ref{LAGSSSG} where we propose a generalisation of the approach of~\cite{BPS}. It shows that the Lagrangian formulation is not unique in general
Altogether, the results of Section~\ref{SSSG} provide the explicit relationship between the equations of motion of the sigma model, the corresponding SSSG equations, and the equations of motion of the gWZW actions that provide their Lagrangian formulation. This is one of the main results of this paper, which is summarised by Figure~\ref{Figura}.

In Section~\ref{Examples}, we illustrate the results of Section~\ref{SSSG} with two examples: The Pohlmeyer reduction of the sigma models with target-space ${\fl C}P^n$, and the reduction of the principal chiral models associated to a compact Lie group $G$, which can be realised as nonlinear sigma models with target-space $G\times G/G_D$. The reduction of the $S^3$ sigma model, which provides the pattern for all the other cases, is discussed in Appendix~\ref{AppCSG}.

In Section~\ref{AdS}, we shall solve the constraints~(\ref{GenConstraint}) for the anti-de Sitter spaces $AdS_n=SO(2,n-1)/SO(1,n-1)$, which are examples of symmetric spaces of indefinite (Lorentzian) signature.
In order to do it, we proof a generalization of the `polar coordinate decomposition' satisfied by symmetric spaces of definite signature. The reduction gives rise to three different basic types of  SSSG equations corresponding to $\mu^2>0$, $\mu^2<0$ and $\mu^2=0$. Their solutions describe bosonic string configurations on ${\fl R}_t\times AdS_n$, $AdS_n\times S^1_\vartheta$ and $AdS_n$, respectively. Finally, Section~\ref{Conclusions} contains our conclusions, and there are two appendices.

\section{Non-linear sigma models on symmetric spaces}
\label{NLSM}

We begin by summarising the construction of (bosonic) nonlinear sigma models with target-space a symmetric space (see~\cite{EFred} for a comprehensive review). 
Let us consider a symmetric space ${\cal M}=F/G$, where $F$ is a connected real Lie group with Lie algebra ${\go f}$, $G$ is a closed subgroup with Lie algebra ${\go g}$, and we have the canonical decomposition~\cite{KOBA,Neill,HEL}
\begin{equation}
{\go f}={\go g}\oplus {\go p},\quad {\rm such\; that} \quad [{\go g},{\go g}]\subset {\go g} , 
\quad [{\go g},{\go p}]\subset 
{\go p} , 
\quad [{\go p},{\go p}]\subset {\go g}\>.
\label{SSalgebra}
\end{equation}
Here, $F=I_0({\cal M})$ is the identity component of the group of isometries of ${\cal M}$ which acts transitively on ${\cal M}$. This means that for each $p,q\in {\cal M}$ there is $f\in F$ such that  $fp=q$ or, equivalently, that ${\cal M}=F\cdot p_0$ for an arbitrary point $p_0\in {\cal M}$. Correspondingly, $G$ is the isotropy group (or little group) of $p_0$; namely, $G=\{g\in F\>:\> gp_0=p_0\}$. We will restrict ourselves to symmetric spaces with $F$ semisimple. Moreover, we will always consider explicit realizations in terms of matrix representations of $F$, and we will assume that the corresponding trace form provides a non-degenerate, $\mathop{\rm Ad}(F)$-invariant, bilinear form on ${\go f}$ such that the decomposition~(\ref{SSalgebra}) is orthogonal.

Let $f=f(\tau,x)$ be a 1+1 dimensional field taking values on a faithful matrix representation of $F$. To formulate the sigma model with target-space ${\cal M}=F/G$, we introduce a gauge field $B_\mu$ on ${\go g}$ and define a covariant derivative $D_\mu f = \partial_\mu f -fB_\mu$
with the property that
\begin{equation}
f\rightarrow fg^{-1}, \quad B_\mu \rightarrow g \bigl(B_\mu  + \partial_\mu\bigr) g^{-1}\quad \Rightarrow\quad
D_\mu f\rightarrow \bigl(D_\mu f\bigr) g^{-1}
\label{Hgauge}
\end{equation}
for any $g=g(\tau,x)$ taking values on $G$. It is also useful to introduce the ${\go f}$-valued current
\begin{equation}
J_\mu = f^{-1} D_\mu f = f^{-1} \partial_\mu f - B_\mu
\label{SMCurrent}
\end{equation}
that is covariant under gauge transformations,
\begin{equation}
J_\mu\rightarrow g J_\mu g^{-1}\>.
\label{GaugeJ}
\end{equation}
Then, if the Lie group $F$ is simple, the nonlinear sigma model is defined by the Lagrangian
\begin{equation}
{\cal L}= -{1\over2\kappa} \mathop{\rm Tr} \bigl(J_\mu J^\mu\bigr),
\label{Lagrangian}
\end{equation}
where $\kappa$ is an overall normalization constant that plays no role in the classical equations of motion. ${\cal L}$ is invariant under the local $G$-symmetry specified by~(\ref{Hgauge}), which exhibits that it is actually defined on the coset $F/G$. In addition, it is also invariant under the global $F$-symmetry $f\rightarrow f_0 f$, for any constant $f_0\in F$. 
Correspondingly, the Lagrangian for $F$ semisimple is a sum of terms like~(\ref{Lagrangian}) with overall normalization factors $\kappa$ that can be different for each simple factor (see~(\ref{LagrangianGen})). 

Without loss of generality, we can restrict ourselves to sigma models defined by Lagragians of the form~(\ref{Lagrangian}). Then, the equation of motion for the field $f$ is
\begin{equation}
D_\mu J^\mu = \partial_\mu J^\mu + [B_\mu, J^\mu]=0
\label{EoM1}
\end{equation}
which, together with the trivial identity $[\partial_\mu + f^{-1}\partial_\mu f, \partial_\mu + f^{-1}\partial_\nu f]=0$, implies
\begin{equation}
D_\mu J_\nu - D_\nu J_\mu + \bigl[J_\mu,J_\nu\bigr] + F_{\mu\nu}=0,
\label{Identity}
\end{equation}
where $F_{\mu\nu}=\partial_\mu B_\nu-\partial_\nu B_\mu+\bigl[B_\mu,B_\nu\bigr]$.
Correspondingly, the equations for the fields $B_\mu$ are
\begin{equation}
J^\mu =0 \quad {\rm on}\quad {\go g}
\label{EoM2}
\end{equation}
which, taking~(\ref{SSalgebra}) into account, force $J^\mu$ to take values in ${\go p}$. Then,~(\ref{EoM1}) and~(\ref{Identity}) split into
\begin{equation}
D_\pm J_\mp =\partial_\pm J_\mp + [B_\pm,J_\mp]=0 \quad ({\rm in}\;\; {\go p})
\quad{\rm and}\quad
[J_+,J_-] + F_{+-}=0 \quad ({\rm in}\;\; {\go g}),
\label{RedEq}
\end{equation}
where we have made use of the light-cone variables $x^\pm ={1\over2}(\tau\pm x)$, and $\partial_\pm = \partial_\tau\pm \partial_x$.
The first equation in~(\ref{RedEq}) implies that 
\begin{equation}
\partial_\pm \mathop{\rm Tr} \bigl(J_\mp^n\bigr)=0 \>,
\label{ChiralJ}
\end{equation}
which provides a set of local chiral densities that display the two-dimensional conformal invariance of the sigma model. In particular, the non-vanishing components of the stress-energy tensor are recovered for $n=2$,
\begin{equation}
T_{++} =-{1\over2\kappa}\mathop{\rm Tr} \bigl(J_+^2\bigr) \quad {\rm and}\quad 
T_{--} =-{1\over2\kappa}\mathop{\rm Tr} \bigl(J_-^2\bigr)\>.
\label{StressTensor2}
\end{equation}
In the following sections, we will consider the class of symmetric space sine-Gordon (SSSG) equations obtained by constraining all the local chiral densities provided by~(\ref{ChiralJ}) to take constant values; namely, the equations obtained by imposing the constraints~(\ref{GenConstraint}),
\begin{equation*}
\mathop{\rm Tr}(J_\pm^n)= \text{\rm constant} \quad \forall \;n\geq2.
\end{equation*}

\section{SSSG equations from sigma models with target-space a symmetric space of definite signature}
\label{SSSG}

The symmetric spaces of definite (positive) signature are characterised by the condition that $G$ and, therefore, ${\go g}$ are compact~\footnote{In the mathematical literature, symmetric spaces of definite signature are usually called Riemannian, while those of indefinite signature are called semi-Riemannian~\cite{Neill}.}.  They have been completely classified by Cartan,
and a thorough survey of their structure and properties can be found in~\cite{HEL}. 
An important result is that any symmetric space ${\cal M}=F/G$ of definite signature with $F$ semisimple can be decomposed as a direct product of symmetric spaces of the following two basic types:
\begin{itemize}
\item[(a)] `Compact type', if ${\go f}$ is compact.
\item[(b)] `Noncompact type', if ${\go f}$ is noncompact and 
${\go f}={\go g}\oplus {\go p}$ is a `Cartan decomposition', which means that the restriction of the trace form to ${\go p}$ is positive definite.\vspace{-4pt}
\end{itemize}
Here we are assuming that the trace form, which is proportional to the Killing form of ${\go f}$, is normalised such that its restriction to any compact subalgebra, and in particular to ${\go g}$, is negative definite. Both types of symmetric spaces are related by the so-called `duality symmetry'
\begin{equation}
{\go f}= {\go g}\oplus {\go p} \longrightarrow {\go f}^\ast={\go g}\oplus i{\go p},
\label{Duality}
\end{equation}
so that if $({\go f},{\go g})$ corresponds to a symmetric space $F/G$ of compact type, then $({\go f}^\ast,{\go g})$ corresponds to a symmetric space $F^\ast/G$ of noncompact type, and the other way around. Examples of symmetric spaces of definite signature of compact and noncompact type are provided by the $n$-spheres
\begin{equation}
S^n=\{(x_1,\ldots,x_{n+1})\>:\> x_1^2+\cdots + x_{n+1}^2 =1\} = SO(n+1)/SO(n) 
\end{equation}
and the $n$-hyperbolic spaces
\begin{equation}
H^n_+=\{(x_1,\ldots,x_{n+1})\>:\> -x_1^2+x_2^2+\cdots + x_{n+1}^2 =-1,\; x_1>0\} = SO(1,n)/SO(n),
\end{equation}
respectively, which are in fact related by the duality symmetry~(\ref{Duality}).
Furthermore, the symmetric spaces of compact type can be decomposed as the direct product of `irreducible' symmetric spaces of two types~\cite{HEL}:
\begin{itemize}
\item[(a.1)] `Type~I', if $F$ is a compact simple Lie group.
\item[(a.2)] `Type~II', if $F/G=G\times G/G_D$, where $G$ is a compact simple Lie group and $G_D$ is the diagonal of the product $G\times G$ (see Section~\ref{ExampleChiralModel}).
\end{itemize}
Since $(G\times G)/G_D$ is trivially isomorphic to $G$, the sigma models with target-space a symmetric space of type~II are just the principal chiral models associated to compact simple Lie groups. The corresponding decomposition of symmetric spaces of noncompact type as the direct product of irreducible symmetric spaces of `type~III' and `type~IV' can be easily deduced from the classification of the compact ones using the duality symmetry~(\ref{Duality}). 

Let us consider a generic symmetric space of definite signature of the form
\begin{equation}
{\cal M} = {\cal M}_-\times  {\cal M}_+,
\end{equation}
where ${\cal M}_-$ and ${\cal M}_+$ are of compact and noncompact type, respectively.
Let $F_{c}$ ($F_{nc}$) be the identity component of the group of isometries of ${\cal M}_-$ (${\cal M}_+$), and $G_-$ ($G_+$)  the isotropy group of an arbitrary point in ${\cal M}_-$ (${\cal M}_+$). Then,
\begin{equation}
{\cal M} = {\cal M}_-\times  {\cal M}_+ = F_{c}/G_- \times F_{nc}/G_+= F/G,
\end{equation}
where $F=F_{c} \times F_{nc}$ and $G=G_-\times G_+$.
By definition, $F_{c}$ is compact and $F_{nc}$ is noncompact. However, since the symmetric space ${\cal M}$ is of definite signature, both $G_-$ and $G_+$ are compact. 
The Lie algebras ${\go f}_c$ of $F_{c}$ and ${\go f}_{nc}$ of $F_{nc}$ admit canonical decompositions of the form~(\ref{SSalgebra}) and, using obvious notation,
${\go g}={\go g}_-\oplus {\go g}_+$ and ${\go p}={\go p}_{c} \oplus {\go p}_{nc}$. By construction, the restriction of the trace form to ${\go p}_{c}$ and ${\go p}_{nc}$ is negative and positive definite, respectively. This provides a non-degenerate, $\mathop{\rm Ad}(F)$-invariant,  bilinear form on ${\go f}$
\begin{equation}
(a,b) =
\begin{cases}
-\mathop{\rm Tr}(ab),& {\rm if}\;\; a,b\in {\go f}_{c}\\
+\mathop{\rm Tr}(ab),& {\rm if}\;\; a,b\in {\go f}_{nc}
\end{cases}
\end{equation}
whose restriction to ${\go p}$ is positive definite, and which extends to the positive definite $F$-invariant metric on ${\cal M}=F/G$. Then, the nonlinear sigma model with target-space ${\cal M}={\cal M}_-\times {\cal M}_+$ is defined by the Lagrangian~\cite{EFII}
\begin{equation}
{\cal L}= {1\over2} \bigl(J_\mu, J^\mu\bigr)=
-{1\over2} \mathop{\rm Tr} \bigl(J_\mu^{(c)} J^{(c)\mu}\bigr)
+{1\over2} \mathop{\rm Tr} \bigl(J_\mu^{(nc)} J^{(nc)\mu}\bigr)
\label{LagrangianGen}
\end{equation}
where $J_\mu^{(c)}$ and $J_\mu^{(nc)}$ are the components of the current~(\ref{SMCurrent}) with respect to the decomposition ${\go f}={\go f}_c\oplus {\go f}_{nc}$. This Lagrangian is a combination of two Lagrangians of the form~(\ref{Lagrangian}) with $\kappa=+1$ and $\kappa=-1$, so that $T_{++}$ and $T_{--}$ are positive definite.
In the following, and without loss of generality, we will assume that ${\cal M}=F/G$ is either of compact or of noncompact type, and that the Lagrangian is of the form~(\ref{Lagrangian}) with $\kappa=+1$ or $-1$, respectively. The generalization of our results to more general cases is straightforward.

When the target-space of the sigma model is a symmetric space of definite signature, the general solution to the constraints~(\ref{GenConstraint}) can be found by using the so-called `polar coordinate decomposition', which is stated as follows~\cite{EFred,HEL}. Let ${\go a}$ be a maximal abelian subspace in ${\go p}$. Then, for any $k\in {\go p}$ there exists $\overline{g}\in G$ such that $\overline{g}^{\>-1} k\> \overline{g}\in {\go a}$. 
The proof of this rather useful property relies on the fact that $G$ is compact. It can be summarised as follows. First of all, it can be proved that ${\go a}$ always contains an element $k_0$ whose centraliser in ${\go p}$ is ${\go a}$; i.e., such that ${\go a} = \{k\in {\go p} \>:\> [k_0,k]=0 \}$.
Then, $g\rightarrow \mathop{\rm Tr}(k g k_0 g^{-1})$ defines a continuous function on the compact group $G$ and, therefore, it takes a minimum for, say, $g=\overline{g}$. For each $T\in {\go g}$, this requires that
\begin{equation}
0= {d\over dx} \mathop{\rm Tr}(k\> \overline{g}\> {\rm e}^{ xT}\> k_0\>  {\rm e}^{ -xT}\> \overline{g}^{-1})\big|_{x=0} =\mathop{\rm Tr} (T\> [k_0, \overline{g}^{-1} k \overline{g}] )\>.
\label{Minimum}
\end{equation}
Since the restriction of the trace form to ${\go g}$ is non-degenerate,~(\ref{Minimum}) implies that $[k_0, \overline{g}^{-1} k \overline{g}]=0$, which ensures that $\overline{g}^{-1} k \overline{g}\in {\go a}$ and completes the proof. 
A more explicit proof specific for $S^n=SO(n+1)/SO(n)$ is given in Appendix~\ref{AppCSG}.

The dimension of the maximal abelian subspaces ${\go a}\subset {\go p}$ defines the rank of the symmetric space. Recall that the rank of the Lie algebras ${\go f}$ and ${\go g}$ is the dimension of their Cartan subalgebras, which are themselves maximal abelian subspaces. Then, taking the decomposition~(\ref{SSalgebra}) into account, it is easy to show that the rank of the symmetric space $F/G$ is bounded as follows
\begin{equation}
\mathop{\rm rank}(F)-\mathop{\rm rank}(G)\leq \mathop{\rm rank}(F/G) \leq \mathop{\rm rank}(F).
\label{RankBound}
\end{equation}
To illustrate these bounds, we have collected the rank of all the symmetric spaces of type~I corresponding to the classical Lie groups in Table~\ref{RankTable}. Notice that there are cases with $\mathop{\rm rank}(F/G) = \mathop{\rm rank}(F)$ where the maximal abelian subspaces of ${\go p}$ are also Cartan subalgebras of ${\go f}$. 
It is also worth noticing that, for symmetric spaces of definite signature, the polar coordinate decomposition ensures that all the maximal abelian subspaces in ${\go p}$ are conjugated under the adjoint action of $G$, a property that is not true for symmetric spaces of indefinite signature like $AdS_n$ (see Section~\ref{AdS}).

\TABLE[ht]{
\begin{tabular}{cccc}
\hline
\\[-10pt]
$F/G$ & $\mathop{\rm rank}(F/G)$ & $\mathop{\rm rank}(F)$ & $\mathop{\rm rank}(G)$ 
\\
\\[-10pt]
\hline
\\[-10pt]
$S^n=SO(n+1)/SO(n)$ & 1 & $\left[n+1\over2\right]$ & $\left[n\over2\right]$\\[5pt]
${\fl C}P^n=SU(n+1)/U(n)$ & 1 & n & n\\[5pt]
$SO(n+m)/SO(n)\times SO(m)$ & $\mathop{\rm min}(n,m)$ & $\left[n+m\over2\right]$ & $\left[n\over2\right]+\left[m\over2\right]$\\[5pt]
$SU(n+m)/S(U(n)\times U(m))$ & $\mathop{\rm min}(n,m)$ & $n+m-1$ & $n+m-1$\\[5pt]
$SU(n)/SO(n)$ & $n-1$ & $n-1$ & $\left[n\over2\right]$\\[5pt]
$SU(2n)/Sp(n)$ & $n-1$ & $2n-1$ & $n$\\[5pt]
$SO(2n)/U(n)$ & $\left[n\over2\right]$ & $n$ & $n$\\[5pt]
$Sp(n)/U(n)$ & $n$ & $n$ & $n$\\[5pt]
$Sp(n+m)/Sp(n)\times Sp(m)$ & $\mathop{\rm min}(n,m)$ & $n+m$ & $n+m$
\\[5pt]
\hline
\\[-5pt]
\end{tabular}
\label{RankTable}
\caption{Rank of the symmetric spaces of type~I corresponding to the classical Lie groups. Here, $\left[q\right]$ denotes the integer part of the rational number $q$. A more complete table can be found in~\cite[Chapter~X]{HEL}.}
}

Now, let us fix ${\go a}$ and denote by ${\go n}$ its orthogonal complement in ${\go p}$, so that ${\go p}={\go a}\oplus {\go n}$ and
the commutation relations~(\ref{SSalgebra}) imply
\begin{equation}
[{\go a},{\go a}]=\{0\}\quad {\rm and}\quad [{\go g},{\go a}]\subset {\go n}\>.
\label{Adecomp}
\end{equation}
Taking~(\ref{EoM2}) into account, we can apply the polar coordinate decomposition to the currents $J_\pm$ so that
\begin{equation}
J_\pm =\overline{g}_\pm c_\pm\> \overline{g}_\pm^{\>-1},
\label{PolarGeneral}
\end{equation}
where $\overline{g}_\pm$ and $c_\pm$ are functions that take values in $G$ and in ${\go a}$, respectively. 
Then, the first equation in~(\ref{RedEq}) becomes
\begin{equation}
\partial_\pm c_\mp = \bigl[\partial_\pm \overline{g}_\mp^{\>-1} \overline{g}_\mp -  \overline{g}_\mp^{\>-1} B_\pm \overline{g}_\mp, \>c_\mp \bigr]
\end{equation}
which, taking~(\ref{Adecomp}) into account, imply that $c_+$ and $c_-$ are chiral,
\begin{equation}
c_+=c_+(x_+)\quad {\rm and}\quad c_-=c_-(x_-)\>.
\end{equation}
Using~(\ref{PolarGeneral}), the chiral densities provided by~(\ref{ChiralJ}) become
\begin{equation}
\mathop{\rm Tr} \bigl(J_\mp^n\bigr)=\mathop{\rm Tr} \bigl(c_\pm^n\bigr),
\end{equation}
which shows that their value is fixed by the value of the components of $c_+$ and $c_-$. This agrees with the results of~\cite{EvansMountain} where it was shown that, for each chirality, the corresponding conserved quantities can be expressed as polynomials in terms of $\mathop{\rm rank}(F/G)$ `primitive' densities, which is  precisely the number of independent components of  $c_+$ and $c_-$.
Therefore, constraining all the chiral densities $\mathop{\rm Tr} \bigl(J_\mp^n\bigr)$ to take constant values is equivalent to constraining the chiral functions $c_+$ and $c_-$ to be constant. 

If $\mathop{\rm rank}(F/G)=1$, it is straightforward to show that this prescription is completely equivalent to the original one implemented by Pohlmeyer. Since $\mathop{\rm dim}{\go a}= 1$, we can write
\begin{equation}
c_+= \mu_+(x_+)\> \Lambda \quad {\rm and}\quad c_-= \mu_-(x_-)\> \Lambda,
\end{equation}
where $\mu_+$ and $\mu_-$ are real (numeric) functions, $\Lambda$ is the only (constant) generator of ${\go a}$ and, since these  symmetric spaces are either of compact or of noncompact type, $\mathop{\rm Tr}  \bigl(\Lambda^2\bigr)\not=0$. Then, according to~(\ref{StressTensor2}),
\begin{equation}
T_{++} = -{1\over2 \kappa}\> \mu_+^2(x_+)  \mathop{\rm Tr} \bigl(\Lambda^2\bigr) \quad {\rm and}\quad 
T_{--} =-{1\over2\kappa}\> \mu_-^2(x_-)  \mathop{\rm Tr} \bigl(\Lambda^2\bigr),
\label{StressTensor3}
\end{equation}
where the value of $\kappa=\pm 1$ is chosen so that $T_{++}$ and $T_{--}$ are always positive, as explained in the paragraph after~(\ref{LagrangianGen}). Therefore, the components of the stress-energy tensor are constant if, and only if, $\mu_+$ and $\mu_-$ are constant, which is obviously equivalent to the claim that $c_+$ and $c_-$ are constant elements of ${\go a}$. 
Since all the maximal abelian subspaces in ${\go p}$ are conjugated under the adjoint action of $G$, in this case the reduction procedure gives rise to only one set of SSSG equations, which are indeed equivalent to the equations of motion of the original sigma model up to a (classical) conformal transformation.

In contrast, if $\mathop{\rm rank}(F/G)>1$ the reduction procedure will give rise to different SSSG equations characterized by the non-equivalent constant values of the primitive chiral densities $\mathop{\rm Tr}(J_\pm^n)$.
Let us take
\begin{equation}
c_+ =\mu_+ \Lambda_+\quad {\rm and}\quad c_- =\mu_- \Lambda_-,
\label{FixElements}
\end{equation}
where $\Lambda_+$ and $\Lambda_-$ are constant elements of ${\go a}$ and $\mu_+,\mu_-$ are real constants.
Then, since either ${\go f}={\go g}\oplus {\go p}$ or ${\go f}^\ast={\go g}\oplus i{\go p}$ is compact, the adjoint actions of $\Lambda_+$ and $\Lambda_-$ in ${\go f}$ can be completely diagonalised, and ${\go f}$ admits the orthogonal decompositions
\begin{equation}
{\go f} = \mathop{\rm Ker} \bigl( {\mathop{\rm Ad}}_{\Lambda_+}\bigr) \oplus \mathop{\rm Im} \bigl( {\mathop{\rm Ad}}_{\Lambda_+}\bigr) \quad{\rm and}\quad
{\go f} = \mathop{\rm Ker} \bigl( {\mathop{\rm Ad}}_{\Lambda_-}\bigr) \oplus \mathop{\rm Im} \bigl( {\mathop{\rm Ad}}_{\Lambda_-}\bigr).
\label{LambdaDec1}
\end{equation}
It is easy to check that they satisfy the commutation relations
\begin{eqnarray}
&&\bigl[\mathop{\rm Ker} \bigl( {\mathop{\rm Ad}}_{\Lambda_\pm}\bigr),
\mathop{\rm Ker} \bigl( {\mathop{\rm Ad}}_{\Lambda_\pm}\bigr)\bigr]
\subset
\mathop{\rm Ker} \bigl( {\mathop{\rm Ad}}_{\Lambda_\pm}\bigr)
\nonumber\\[5pt]
&&
{\rm and}\quad \bigl[\mathop{\rm Ker} \bigl( {\mathop{\rm Ad}}_{\Lambda_\pm}\bigr),
\mathop{\rm Im} \bigl( {\mathop{\rm Ad}}_{\Lambda_\pm}\bigr)\bigr]
\subset
\mathop{\rm Im} \bigl( {\mathop{\rm Ad}}_{\Lambda_\pm}\bigr)\>.
\label{LambdaDec2}
\end{eqnarray}
In the following, we will need the centralisers of $\Lambda_+$ and $\Lambda_-$ in $G$, which are the two Lie groups
\begin{equation}
H^{(\pm)} =\bigl\{ g\in G \>:\> g^{-1}\Lambda_\pm g= \Lambda_\pm \bigr\},
\end{equation}
with Lie algebras
\begin{equation}
{\go h}_\pm = \mathop{\rm Ker} \bigl( {\mathop{\rm Ad}}_{\Lambda_\pm}\bigr)\cap {\go g}\>.
\end{equation}
It is worth noticing that, in general, $H^{(+)}\not= H^{(-)}$ and ${\go h}_+\not={\go h}_-$.

The explicit formulation of the reduced model is obtained by imposing a particular gauge-fixing condition to the equations of motion of the sigma model subjected to the constraints~(\ref{FixElements}).
Namely,~(\ref{GaugeJ}) and~(\ref{PolarGeneral}) enable the so-called `partial reduction' gauge condition~\cite{EFred}
\begin{equation}
J_+ = \mu_+ \Lambda_+\quad {\rm and}\quad J_- = \mu_- \gamma^{-1} \Lambda_- \gamma,
\label{RedCurrents}
\end{equation}
where $\gamma=\overline{g}_-^{\>-1} \overline{g}_+$ takes values in $G$. Then, the first two equations in~(\ref{RedEq}), $D_\pm J_\mp=0$, become
\begin{equation}
[B_-,\Lambda_+]=0\quad {\rm and}\quad
[B_+ - \gamma^{-1}\partial_+\gamma , \gamma^{-1}\Lambda_-\gamma]=0,
\end{equation}
whose general solution is
\begin{equation}
B_- = A_-^{(R)} \in {\go h}_+ \quad {\rm and}\quad
B_+ = \gamma^{-1} \partial_+ \gamma + \gamma^{-1} A_+^{(L)} \gamma, \quad{\rm with} \quad A_+^{(L)}\in{\go h}_-\>.
\label{RedGauge}
\end{equation}
The condition~(\ref{RedCurrents}) does not fix the gauge symmetry~(\ref{Hgauge}) completely, and the residual gauge transformations correspond to $\gamma\rightarrow \gamma h_+^{-1}$, with
$h_+\in H^{(+)}$. Moreover,~(\ref{RedCurrents}) is also explicitly invariant under $\gamma\rightarrow  h_-\gamma$, with $h_-\in H^{(-)}$. 
Taking~(\ref{Hgauge}) also into account, all these gauge transformations can be summarised as follows:
\begin{equation}
\gamma \rightarrow h_-\> \gamma\> h_+^{-1}, \quad A_-^{(R)}\rightarrow h_+\bigl( A_-^{(R)}  + \partial_-\bigr) h_+^{-1}, \quad{\rm and}\quad  A_+^{(L)}\rightarrow h_- \bigl(A_+^{(L)}  + \partial_+\bigr) h_-^{-1}\>.
\label{ESSSG1}
\end{equation}
The third equation in~(\ref{RedEq}) can be written as a zero-curvature condition for $\gamma$,
\begin{equation}
\bigl[\>\partial_+ + \gamma^{-1}\partial_+\gamma + \gamma^{-1} A_+^{(L)} \gamma + z \mu_+\> \Lambda_+,\> \partial_- + A_-^{(R)} + z^{-1} \mu_- \>\gamma^{-1} \Lambda_-\gamma \>\bigl]=0
\label{ESSSG2}
\end{equation}
where $z$ is a spectral parameter.
This equation, subjected to the gauge symmetry~(\ref{ESSSG1}), provides the most general form of the SSSG equations specified by $(F/G,\Lambda_+,\Lambda_-)$. 
Actually, it is the integrability condition required to reconstruct the field $f=f(\tau,x)$ corresponding to the currents~(\ref{RedCurrents}) and the gauge fields~(\ref{RedGauge}). Namely, using~(\ref{SMCurrent}),~$f$ is the solution to the auxiliary linear problem
\begin{eqnarray}
&&
\partial_+ f^{-1} = -\left(\gamma^{-1}\partial_+\gamma + \gamma^{-1} A_+^{(L)} \gamma +  \mu_+\> \Lambda_+\right) f^{-1}\nonumber\\[5pt]
&&
\partial_- f^{-1} = -\left(A_-^{(R)} +  \mu_- \>\gamma^{-1} \Lambda_-\gamma\right) f^{-1}
\end{eqnarray}
whose integrability condition is~(\ref{ESSSG2}). It has a unique solution 
once the initial condition $f_0=f(\tau_0,x_0)$ is fixed.
The zero-curvature condition~(\ref{ESSSG2}) subjected to~(\ref{ESSSG1}) exhibits classically integrability and two-dimen\-sional Lorentz invariance. Moreover, it shows that the model is naturally defined on the left-right asymmetric coset
\begin{equation}
G/H^{(-)}_L\times H^{(+)}_R = G/\bigl[\>\gamma\sim h_-\> \gamma\> h_+^{-1};\; \gamma\in G, h_- \in H^{(-)}, \>h_+ \in H^{(+)}\>\bigr],
\end{equation}
which consists of orbits
under the action of $H^{(-)}_L\times H^{(+)}_R$ on $G$.

An interesting case occurs when 
the symmetric space is of maximal rank, which means that $\mathop{\rm rank}(F/G)=\mathop{\rm rank}(F)$. Then, the maximal abelian subspaces in ${\go p}$ are Cartan subalgebras of ${\go f}$, and it is possible to choose $\Lambda_+$ and $\Lambda_-$ such that $H^{(+)}= H^{(-)}=\{1\}$. The corresponding SSSG equations are defined on the group manifold $G$, and some features of the integrable models related to them have been discussed in~\cite{SSG}.

\subsection{Connection with the non-abelian affine Toda (NAAT) equations}
\label{NAAT}

The SSSG equations are usually written as a system of NAAT equations, which amounts to fixing the gauge symmetry~(\ref{ESSSG1}) as follows. First of all, we shall write the zero-curvature condition~(\ref{ESSSG2}) as
\begin{equation}
\bigl[\>\partial_+ + \gamma^{-1}\partial_+\gamma + \gamma^{-1} A_+^{(L)} \gamma,\> \partial_- + A_-^{(R)}  \>\bigl]
+\mu_+\mu_- \bigl[\> \Lambda_+,\> \gamma^{-1} \Lambda_-\gamma 
\>\bigl]=0.
\label{SSSG3}
\end{equation}
Taking~(\ref{LambdaDec1}) and~(\ref{LambdaDec2}) into account, it leads to
\begin{eqnarray}
&&\bigl[\>\partial_+ + {\mathop{\rm P}}_{{\go h}_+}(\gamma^{-1}\partial_+\gamma + \gamma^{-1} A_+^{(L)} \gamma),\> \partial_- + A_-^{(R)}  \>\bigl]
=0 \nonumber\\[5pt]
&&
{\rm and}\quad
\bigl[\>\partial_+ + A_+^{(L)} ,\> \partial_- + {\mathop{\rm P}}_{{\go h}_-}(-\partial_-\gamma\gamma^{-1} + \gamma A_-^{(R)} \gamma^{-1})  \>\bigl]
=0,
\label{SSSG4}
\end{eqnarray}
where ${\mathop{\rm P}}_{{\go h}_+}$ and ${\mathop{\rm P}}_{{\go h}_-}$ are the orthogonal projectors on ${\go h}_+$ and ${\go h}_-$, respectively. These equations enable the gauge-fixing conditions
\begin{equation}
A_-^{(R)}=A_+^{(L)}={\mathop{\rm P}}_{{\go h}_+}(\gamma^{-1}\partial_+\gamma)= {\mathop{\rm P}}_{{\go h}_-}(\partial_-\gamma\gamma^{-1})
=0
\label{SSSGLSgauge}
\end{equation}
and, in this gauge, the SSSG equations~(\ref{SSSG3}) and~(\ref{SSSG4}) reduce to
\begin{eqnarray}
&&
\partial_-(\gamma^{-1}\partial_+\gamma)=\mu_+\mu_- \bigl[\> \Lambda_+,\> \gamma^{-1} \Lambda_-\gamma 
\>\bigl]\nonumber\\[5pt]
&&
{\rm and}\quad {\mathop{\rm P}}_{{\go h}_+}(\gamma^{-1}\partial_+\gamma)= {\mathop{\rm P}}_{{\go h}_-}(\partial_-\gamma\gamma^{-1})
=0,
\label{ENAAT}
\end{eqnarray}
which constitute a system of NAAT equations associated to the ${\fl Z}_2$-gradation of ${\go f}$ given by~(\ref{SSalgebra})~\cite{NAATls,NAATtau,NAATkir} (see also~\cite{HMP,FMSG,HSG}).~\footnote{For earlier references on non-abelian generalizations of the sine-Gordon see~\cite{BUDAGOV}.}

\subsection{Lagrangian formulation}
\label{LAGSSSG}

The Lagrangian formulation of the SSSG equations have been a long-standing problem until Bakas, Park and Shin proposed to identify them with the equations of motion of a gauged Wess-Zumino-Witten (gWZW) action with a potential term~\cite{BPS} (see also~\cite{Parkold,HMP,FMSG,HSG}). However, the relationship between the original sigma model degrees of freedom and those of the modified gWZW action was not clarified in that article which,
in particular, hides that the Lagrangian formulation also involves a particular choice of the gauge fixing conditions~\cite{GTseytlin1}.

One of the reasons behind that omission is that~\cite{BPS} starts directly with  the formulation of the SSSG equations in terms of NAAT equations like~(\ref{ENAAT}).
Actually, it is important to point out that not all the NAAT equations considered in that paper correspond to reduced symmetric space sigma models. To be precise, recall that although the equations considered in~\cite{BPS} are also specified by the data $(F/G,\Lambda_+,\Lambda_-)$, they are of the form
\begin{eqnarray}
&&
\partial_-(\gamma^{-1}\partial_+\gamma)=\mu^2 \bigl[\> \Lambda_+,\> \gamma^{-1} \Lambda_-\gamma 
\>\bigl]\nonumber\\[5pt]
&&
{\rm along\; with}\quad
{\mathop{\rm P}}_{{\go h}}(\gamma^{-1}\partial_+\gamma)= {\mathop{\rm P}}_{{\go h}}(\partial_-\gamma\gamma^{-1})
=0,
\label{NAATBPS}
\end{eqnarray}
where ${\go h}$ is the simultaneous centraliser of $\Lambda_+$ and $\Lambda_-$ in the Lie algebra ${\go g}$; {\em i.e.},
\begin{equation}
{\go h}=C_{\go{g}}(\Lambda_+,\Lambda_-)=\{ r\in {\go g}\>:\> [r,\Lambda_+]=0=[r,\Lambda_-] \}\>.
\end{equation}
Clearly, $C_{\go{g}}(\Lambda_+,\Lambda_-)={\go h}_+\cap {\go h}_-$ and, by comparison with~(\ref{ENAAT}), it is straightforward to notice that the system of NAAT equations~(\ref{NAATBPS}) describes a reduced symmetric space sigma model only if ${\go h}_+= {\go h}_-$. In other words, the construction of~\cite{BPS} provides a Lagrangian formulation only for the SSSG equations corresponding to $\Lambda_+$ and $\Lambda_-$ such that $H^{(+)}=H^{(-)}$.

Nevertheless, the results of~\cite{BPS} can be generalised to a more general class of models where $H^{(+)}$ and $H^{(-)}$ are isomorphic, but they can be allowed to be different.
To be specific, we shall consider the class of SSSG equations associated to $\Lambda_+$ and $\Lambda_-$ such that 
\begin{equation}
H^{(+)}=\epsilon_R(H)\quad {\rm and}\quad H^{(-)}=\epsilon_L(H),
\label{Embed}
\end{equation}
where $H$ is a Lie group with Lie algebra ${\go h}$, and $\epsilon_{R/L}: H\rightarrow G$ are two group homomorphisms that descend to embeddings of the corresponding Lie algebras. We will also require that
these homomorphisms are `anomaly free', which simply means that~\cite{gWZW1,gWZW2,gWZW3,AsymgWZW1,AsymgWZW2}
\begin{equation}
\mathop{\rm Tr}\bigl(\epsilon_L(a)\> \epsilon_L(b) \bigr)
=\mathop{\rm Tr}\bigl(\epsilon_R(a)\> \epsilon_R(b) \bigr)
, \quad \forall\> a,b\in{\go h}\>.
\label{AnomalyFree}
\end{equation}
An important observation is that the choice of $\epsilon_L$ and $\epsilon_R$ in~(\ref{Embed}) is not unique and, therefore, the Lagrangian formulation will not be unique in general either.

Provided that the Lie groups $H^{(+)}$ and $H^{(-)}$ satisfy~(\ref{Embed}), it is easy to show that the system of NAAT equations~(\ref{ENAAT}) admits a Lagrangian formulation in terms of the action
\begin{equation}
S\bigl[\gamma,{\cal A}_\pm\bigr]=S_{\rm gWZW}\bigl[\gamma,{\cal A}_\pm\bigr]-{\mu_+\mu_-\over 2\pi} \int d^2 x\> \mathop{\rm Tr}\bigl(\Lambda_+ \gamma^{-1}\Lambda_- \gamma\bigr)\>.
\label{SSSGAction}
\end{equation}
Here, $\gamma\in G$, ${\cal A}_\pm \in{\go h}$, and
\begin{eqnarray}
&&
S_{\rm gWZW}\bigl[\gamma,{\cal A}_\pm\bigr]=
S_{\rm WZW}\bigl[\gamma\bigr] + {1\over 2\pi} \int d^2 x \mathop{\rm Tr}\Bigl(-\epsilon_L({\cal A}_+)\> \partial_-\gamma \gamma^{-1}+
\nonumber\\[5pt]
&&
\hspace{3.5cm}+\epsilon_R({\cal A}_-)\> \gamma^{-1} \partial_+\gamma+\gamma^{-1}\>\epsilon_L({\cal A}_+)\> \gamma\> \epsilon_R({\cal A}_-) - {\cal A}_+\> {\cal A}_-\Bigr)\>
\label{Action}
\end{eqnarray}
is the gWZW action associated to the asymmetric coset~\cite{AsymgWZW1,AsymgWZW2}~\footnote{To be precise, the coset model is asymmetric only if $\epsilon_L\not=\epsilon_R$. Although we will not display the dependence of the coset $G/H$ on the choice of $\epsilon_L$ and $\epsilon_R$, it is important to stress that its geometry is very sensitive to it.}
$$
G/H= G/\bigl[\>\gamma\sim \epsilon_L(h)\> \gamma\> \epsilon_R(h^{-1});\; \gamma\in G,\; h \in H\>\bigr].
$$
The action~(\ref{SSSGAction}) is invariant under
\begin{equation}
\gamma\rightarrow \epsilon_L(h)\> \gamma\> \epsilon_R(h^{-1})\quad {\rm and}\quad {\cal A}_\pm \rightarrow  h\bigl({\cal A}_\pm  +\partial_\pm\bigr)h^{-1},\quad \forall\; h\in H\>.
\label{Gauge2}
\end{equation}
The equation of motion for the field $\gamma$ can be written as the zero-curvature condition
\begin{equation}
\bigl[\>\partial_+ + \gamma^{-1}\partial_+\gamma + \gamma^{-1} \epsilon_L({\cal A}_+) \gamma + z \mu_+\> \Lambda_+,\> \partial_- + \epsilon_R({\cal A}_-) + z^{-1} \mu_- \>\gamma^{-1} \Lambda_-\gamma \>\bigl]=0\>,
\label{EoMgWZW1}
\end{equation}
and the equations for the fields ${\cal A}_\pm$ are
\begin{eqnarray}
&&
{\mathop{\rm P}}_{{\go h}_-}\Bigl(-\partial_-\gamma \gamma^{-1} + \gamma \epsilon_R({\cal A}_-)\gamma^{-1}\Bigr) = \epsilon_L({\cal A}_-),\nonumber\\[5pt]
&&
{\rm and}\quad
{\mathop{\rm P}}_{{\go h}_+}\Bigl(\gamma^{-1}\partial_+\gamma  + \gamma^{-1} \epsilon_L({\cal A}_+)\gamma\Bigr) = \epsilon_R({\cal A}_+),
\label{EoMgWZW2}
\end{eqnarray}
where we remind that ${\go h}_-=\epsilon_L({\go h})$ and ${\go h}_+=\epsilon_R({\go h})$. 
Then, the connection with the NAAT equations is recovered as follows. Using~~(\ref{LambdaDec1}
),~(\ref{LambdaDec2}) and~(\ref{EoMgWZW2}), the ${\go h}_+=\epsilon_R({\go h})$ component of~(\ref{EoMgWZW1}) implies that the ${\go h}$-connection ${\cal A}_\pm$ is flat,
\begin{equation}
\bigl[ \partial_+ +{\cal A}_+ ,\> \partial_- +{\cal A}_-\bigr]=0,
\label{gWZWFlat}
\end{equation}
which enables the gauge condition
\begin{equation}
{\cal A}_+={\cal A}_-=0.
\label{GaugeNAAT}
\end{equation}
In this gauge,~(\ref{EoMgWZW1}) and~(\ref{EoMgWZW2}) become just the NAAT equations~(\ref{ENAAT}).

However, our main interest here is to establish the relationship between the degrees of freedom of the original sigma model and those of the action~(\ref{SSSGAction}). It follows from the comparison of~(\ref{ESSSG1}) and (\ref{ESSSG2}) with~(\ref{Gauge2}) and (\ref{EoMgWZW1}), respectively.
Indeed, notice that with the identifications
\begin{equation}
A_+^{(L)}= \epsilon_L({\cal A}_+)\quad {\rm and}\quad
A_-^{(R)}= \epsilon_R({\cal A}_-)
\label{FinalGauge}
\end{equation}
the zero-curvature conditions~(\ref{ESSSG2}) and~(\ref{EoMgWZW1}) coincide. Correspondingly, the interpretation of the constraints~(\ref{EoMgWZW2}) was clarified by Grigoriev and Tseytlin in~\cite{GTseytlin1}. They are gauge conditions that partly fix the symmetry under $H^{(-)}_L\times H^{(+)}_R$ given by~(\ref{ESSSG1}), so that the residual gauge transformations correspond to
\begin{equation}
h_+=\epsilon_R(h), \quad h_-=\epsilon_L(h), \quad h\in H
\end{equation}
or, equivalently, to~(\ref{Gauge2}). 
The consistency of
this interpretation of the constraints~(\ref{EoMgWZW2}) for symmetric spaces of definite signature will be demonstrated in Appendix~\ref{AppGauge}.

It is worth noticing that the interpretation of~(\ref{EoMgWZW2}) as gauge conditions implies that the different Lagrangian formulations provided by different choices of $\epsilon_L$ and $\epsilon_R$ are related by gauge transformations. From the point of view of the relevant gWZW actions with a potential term, those gauge transformations should correspond to target-space duality symmetries similar to those discussed in~\cite{HSGphases}. 
We will explicitly check that this is so in the reduction of the ${\fl C}P^2$ and $S^3$ sigma models discussed in Section~\ref{ExampleCPn} and
Appendix~\ref{AppCSG}, respectively.

Altogether, the results of this section provide the explicit map between the solutions $(f,B_\mu)$ to the equations of motion~(\ref{ESSSG2}) of the reduced sigma model specified by $(F/G,\Lambda_+,\Lambda_-)$, and the solutions $(\gamma,{\cal A}_\mu)$ to the equations of motion of the $G/H$ gWZW action with a potential term~(\ref{SSSGAction}). 
Its form, which constitutes one of the main results of this paper, is summarised by Figure~\ref{Figura}.

\newcommand{\sts}{\footnotesize}
\setlength{\unitlength}{1mm}
\newsavebox{\Reduction}
\sbox{\Reduction}
{\begin{picture}(120,54)
\put(0,0){\line(1,0){34}}
\put(0,0){\line(0,-1){15}}
\put(2,-5){$F/G$ sigma model}
\put(10,-11){$(f,B_\pm)$}
\put(0,-15){\line(1,0){34}}
\put(34,-15){\line(0,1){15}}
\put(36,-7){\vector(1,0){41}}
\put(41,0){$\Lambda_+,\Lambda_-$ reduction}
\put(39,-5){eqs.~(\ref{RedCurrents}) and~(\ref{RedGauge})}
\put(79,0){\line(1,0){34}}
\put(79,0){\line(0,-1){16}}
\put(79,-16){\line(1,0){34}}
\put(113,-16){\line(0,1){16}}
\put(81,-5){$G/(H_L^{(-)}\otimes H_R^{(+)})$}
\put(84,-13){$(\gamma,A_+^{(L)},A_-^{(R)})$}
\put(90,-18){\vector(0,-1){20}}
\put(82,-39){\line(1,0){28}}
\put(82,-39){\line(0,-1){15}}
\put(82,-54){\line(1,0){28}}
\put(110,-54){\line(0,1){15}}
\put(84,-44){$G/H\;\;$gWZW}
\put(90,-50){$(\gamma,{\cal A}_\pm)$}
\put(93,-25){Eqs.~(\ref{Embed}),~(\ref{FinalGauge}) and },
\put(93,-31){gauge fixing~(\ref{EoMgWZW2})}
\put(79,-46){\vector(-1,0){42}}
\put(0,-39){\line(1,0){34}}
\put(0,-39){\line(0,-1){15}}
\put(3,-44){NAAT equation}
\put(13,-50){$(\gamma)$}
\put(0,-54){\line(1,0){34}}
\put(34,-54){\line(0,1){15}}
\put(46,-51){Gauge fixing}
\put(49,-56){eq.~(\ref{GaugeNAAT})}
\put(85,-18){\line(0,-1){11}}
\put(85,-29){\line(-1,0){67}}
\put(18,-29){\vector(0,-1){8}}
\put(46,-22){Gauge fixing}
\put(49,-27){eq.~(\ref{SSSGLSgauge})}
\end{picture}}
\FIGURE[ht]{
\begin{picture}(120,58)
(5,-72)
\put(10,-15){\usebox{\Reduction}}
\end{picture}
\label{Figura}
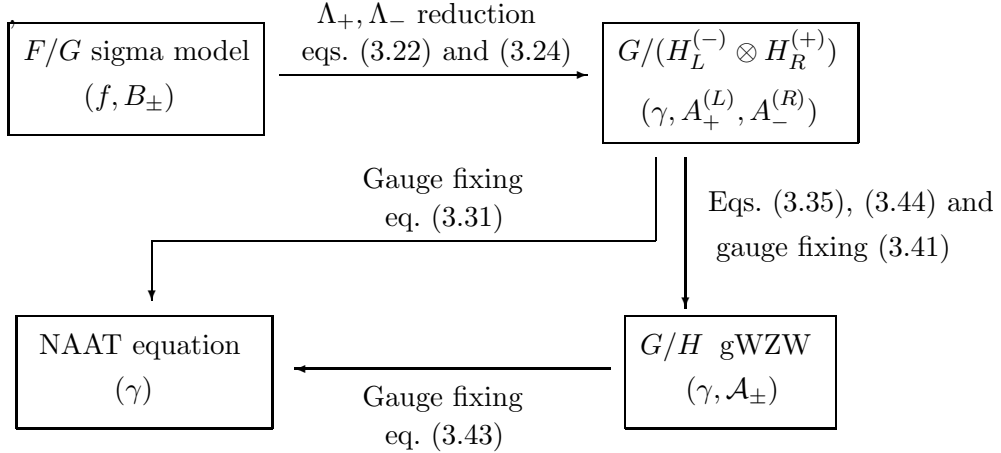
\caption{A schematic summary of the map between the solutions $(f,B_\mu)$ to the equations of motion~(\ref{ESSSG2}) of the reduced sigma model specified by $(F/G,\Lambda_+,\Lambda_-)$, the solutions $(\gamma,{\cal A}_\mu)$ to the equations of motion of the $G/H$ gWZW action with a potential term~(\ref{SSSGAction}), and the solutions $(\gamma)$ to the NAAT equations~(\ref{ENAAT}).}
}

\section{Examples}
\label{Examples}

\subsection{Pohlmeyer reduction of ${\fl C}P^n$ sigma models: Bosonic strings on ${\fl R}_t\times {\fl C}P^n$}
\label{ExampleCPn}

We shall illustrate the main features of the construction presented in the previous section with the Pohlmeyer reduction of the ${\fl C}P^n$ nonlinear sigma model.  The solutions to the resulting SSSG equations describe bosonic string configurations on ${\fl R}_t\times {\fl C}P^n$, which could be useful in the study of the recently proposed
duality between superstring theory on $AdS_4 \times {\fl C}P^3$ and
$N = 6$ super Chern-Simons theory~\cite{NEW}.
The reduction of the ${\fl C}P^n$ nonlinear sigma model has been discussed long ago by Eichenhrr and Honerkamp in~\cite{ECPn,EHCPn}, and its Lagrangian formulation in terms of a gWZW action with a potential term was proposed by Bakas, Park and Shin in~\cite{BPS}. Our construction clarifies the relationship between the degrees of freedom of that action and those of the original sigma model. In particular, we shall discuss in detail the case of ${\fl C}P^2$, which admits two different Lagrangians and illustrates the non-uniqueness of the Lagrangian formulation pointed out in Section~\ref{LAGSSSG}~\footnote{The explicit form of the SSSG equations corresponding to ${\fl C}P^3$ has been recently worked out in~\cite{Rashkov}, in the context of the study of superstring theory on $AdS_4 \times {\fl C}P^3$.}

The $2n$-dimensional complex projective space
\begin{eqnarray}
{\fl C}P^n &=&{\{(z_1,\ldots,z_{n+1})\in {\fl C}^{n+1}\}/ [\>(z_1,\ldots,z_{n+1})\sim \lambda (z_1,\ldots,z_{n+1});\> \lambda \in{\fl C}\not=0\>]}\\[7pt]
&=&
SU(n+1)/U(n)
\label{CPnDefinition}
\end{eqnarray}
is a compact symmetric space of (definite signature and) type~I. Using the $(n+1)\times (n+1)$ matrix representation of $SU(n+1)$, we can choose the embedding of $U(n)=SU(n)\times U(1)$ into $SU(n+1)$ to be
\begin{equation}
(M,{\rm e}^{\> i\phi})\;\rightarrow\;
\left(\begin{array}{cccc}
{\rm e}^{i n\phi} & 0&\cdots&0 \\
0&&& \\
\vdots & &{\rm e}^{-i\phi}\> M&\\
0&&&
\end{array}
\right),
\end{equation}
which corresponds to the isotropy group of the point $(1,0,\ldots,0)$. Then, the form of the elements $r\in {\go g}$ and $k\in{\go p}$ in~(\ref{SSalgebra}) is
\begin{equation}
r=
\left(\begin{array}{cccc}
i n \phi\; & 0 & \cdots & 0 \\
0\; & & & \\
\vdots\; & &-i \phi\>{\fl I}+ \cal{M}& \\
0\; & &  & 
\end{array}
\right)
\quad {\rm and}\quad
k=
\left(\begin{array}{cccc}
0 &v_1&\cdots&v_n \\
-v_1^\ast & 0&\cdots&0\\
\vdots&\vdots&\cdots&\vdots\\
-v_n^\ast & 0&\cdots&0
\end{array}
\right),
\end{equation}
where~${\cal M}=-{\cal M}^\dagger$ denotes a $n\times n$ anti-Hermitian matrix, $(v_1,\ldots,v_n)$ is a complex $n$-dimensional file vector, and $\phi$ is real. It is straightforward to check that this decomposition is orthogonal with respect to the trace form and, moreover, that the rank of ${\fl C}P^n$ is~1.

Since $\mathop{\rm rank}({\fl C}P^n)=1$, the polar coordinate decomposition ensures that any choice of $\Lambda_+=\Lambda_-\in{\go p}$ in~(\ref{RedCurrents}) gives rise to the same set of SSSG equations;
we shall use
\begin{equation}
\Lambda_+=\Lambda_-=
\left(\begin{array}{ccccc}
0 &1&0&\cdots&0 \\
-1 & 0&0&\cdots&0\\
0&0&0&\cdots&0\\
\vdots&\vdots&\vdots&\cdots&\vdots\\
0&0&0&\cdots&0
\end{array}
\right).
\label{LambdaCPn}
\end{equation}
Then, the elements in the centraliser of $\Lambda_+=\Lambda_-$ in $U(n)$ are of the form
\begin{equation}
\left(\begin{array}{ccccc}
{\rm e}^{i(n-1)\phi}&0&0&\cdots&0 \\
0& {\rm e}^{i(n-1)\phi}&0&\cdots&0\\
0&0&&&\\
\vdots&\vdots&&{\rm e}^{-2i\phi} M&\\
0&0&&&
\end{array}
\right),
\end{equation}
where $\phi$ is real and $M\in SU(n-1)$. Therefore,\begin{equation}
H^{(+)}=H^{(-)} \simeq U(n-1)
\end{equation}
and, following~\cite{BPS} and Section~\ref{LAGSSSG}, the resulting SSSG equations admit a (non-unique) Lagrangian formulation provided by a gWZW action associated to the coset 
$U(n)/U(n-1)$
with a potential term. It follows the pattern summarised by figure~\ref{Figura}.

The simplest case is the reduction of the ${\fl C}P^1$ model, which is identical to the $S^2$ sigma model. Then, $H^{(+)}=H^{(-)}$ are trivial, and the corresponding SSSG equation is the sine-Gordon equation.

The first case with non-trivial gauge groups is provided by the reduction of the ${\fl C}P^2=SU(3)/U(2)$ sigma model, which can be used to describe bosonic string configurations on ${\fl R}_t\times {\fl C}P^2$. In this case, 
\begin{equation}
H^{(+)}=H^{(-)}=\{e^{\>x T}\>:\> x\in{\fl R}\}\simeq U(1),
\quad {\rm with}\quad
T=
\left(\begin{array}{ccc}
i & 0 &0 \\
0 & i &0\\
0& 0 & -2i
\end{array}
\right),
\label{CP2algebra}
\end{equation}
which motivates the parameterisation
\begin{equation}
\gamma= {\rm e}^{\>\alpha \>T}\;
\left(\begin{array}{ccc}
1&0&0\\
0&\cos \theta\>{\rm e}^{i \varphi}&\sin \theta\\
0&-\sin \theta\;\;&\cos \theta\>{\rm e}^{-i \varphi}
\end{array}\right)\;
{\rm e}^{-\beta \>T},
\quad
A_-^{(R)}=a_- \>T
\quad{\rm and}\quad
A_+^{(L)}=a_+ \>T
\end{equation}
in terms of six real fields $\alpha$, $\beta$, $\theta$, $\varphi$, $a_+$ and $a_-$. Then, the $U(1)_L\times U(1)_R$ gauge transformations~(\ref{ESSSG1}) read
\begin{equation}
\alpha\rightarrow \alpha +\rho_-, \quad
\beta\rightarrow \alpha +\rho_+\quad
{\rm and}\quad a_\pm \rightarrow a_\pm -\partial_\pm \rho_\mp,
\label{CPgauge}
\end{equation}
where $h_\pm={\rm e}^{\>\rho_\pm\> T}$, and the gauge invariance of the zero-curvature equations~(\ref{ESSSG2}) ensures that they can be written in terms of the four gauge invariant fields
\begin{equation}
\theta, \quad \varphi, \quad b_+ = a_+ + \partial_+\alpha, \quad{\rm and}\quad
b_- = a_- + \partial_-\beta.
\label{GaugeTransCP2}
\end{equation}
In this case Pohlmeyer reduction gives rise to precisely four partial differential equations. Two of them are the continuity equations
\begin{eqnarray}
&&
4\partial_+b_- - \partial_-\left[ \bigl(1+3\cos(2\theta)\bigr)b_+ + 2 \cos^2 \theta\> \partial_+ \varphi\right]=0
\label{CE1}\\[5pt]
&&
\quad{\rm and}\quad
\partial_+\left[ \bigl(1+3\cos(2\theta)\bigr)b_- - 2 \cos^2 \theta\> \partial_- \varphi\right]- 4\partial_-b_+=0,
\label{CE2}
\end{eqnarray}
which correspond to~(\ref{SSSG4}). They can be used to write $b_+$ and $b_-$ in terms of a new field, which can be done in two different ways that lead to different Lagrangians.

Let us consider first the subtraction of~(\ref{CE2}) from~(\ref{CE1}), which reads
\begin{equation}
\partial_+\bigl[ 3\sin^2 \theta\>b_- +\cos^2\theta\> \partial_-\varphi)\bigr] -
\partial_-\bigl[-3\sin^2 \theta\>b_+ +\cos^2\theta\> \partial_+\varphi)\bigr]=0.
\label{CE3}
\end{equation}
It provides the integrability conditions for
\begin{equation}
b_\pm =\pm{1\over 3\sin^2 \theta} \left(\partial_\pm \psi + \cos^2\theta\>\partial_\pm\varphi\right),
\label{Newfield1}
\end{equation}
where $\psi$ is the new real field. Writing now~(\ref{CE1}) in terms of $\psi$, and after some trivial algebra, we get the conservation law
\begin{equation}
\partial_\mu\left[{1\over2}\partial^\mu\psi + 2\cot^2\theta\> \partial^\mu (\psi+\varphi)\right] =0.
\label{CPnEoM1}
\end{equation}
Furthermore, using~(\ref{Newfield1}), the other two equations provided by Pohlmeyer reduction are
\begin{eqnarray}
&&
\partial_\mu \partial^\mu \psi- 4\mu_+\mu_- \cos\theta\sin\varphi=0
\label{CPnEoM2}\\[5pt]
&&
\partial_\mu \partial^\mu \theta + {\cos\theta\over \sin^3\theta}\>\partial_\mu (\psi+\varphi) \partial^\mu(\psi+\varphi) + \mu_+\mu_- \sin\theta\cos\varphi=0,
\label{CPnEoM3}
\end{eqnarray}
and it is straightforward to check that~(\ref{CPnEoM1}--\ref{CPnEoM3}) are the equations of motion of
\begin{equation}
{\cal L}=\partial_\mu\theta \partial^\mu\theta + 
{1\over4}\>\partial_\mu\psi\partial^\mu\psi+\cot^2\theta\>\partial_\mu(\psi+\varphi) \partial^\mu(\psi+\varphi)+ 2\mu_+\mu_- \cos\theta \cos\varphi,
\label{LagrangianEH}
\end{equation}
which is the Lagrangian originally obtained by Eichenherr and Honerkamp in~\cite{EHCPn}.

However, there is an alternative way to write $b_+$ and $b_-$ in terms of a additional real field. Consider now the sum of~(\ref{CE1}) and~(\ref{CE2}), which reads
\begin{equation}
\partial_+\bigl[(1+3\cos^2\theta)b_- -\cos^2\theta\>\partial_-\varphi]-
\partial_-\bigl[(1+3\cos^2\theta)b_+ +\cos^2\theta\>\partial_-\varphi]=0.
\end{equation}
Then,
\begin{equation}
b_\pm =
{1\over 1+3\cos^2\theta} \left(\partial_\pm \widetilde\psi \mp \cos^2\theta\> \partial_\pm\varphi\right)
\label{Newfield2}
\end{equation}
where $\widetilde{\psi}$ is another real field. Writing now~(\ref{CE1}) in terms of $\widetilde{\psi}$, we get the new conservation law
\begin{equation}
\partial_\mu \left({1\over 1+4\cot^2\theta}\Bigl(3\>\partial^\mu \widetilde\psi -\>\epsilon^{\mu\nu}\> \partial_\nu\varphi \Bigr) \right)=0.
\end{equation}
where $\epsilon_{01}=\epsilon^{10}=+1$.
This conservation law and the two other equations provided by Pohlmeyer reduction written in terms of $\widetilde{\psi}$ turn out to be the equations of motion of
\begin{eqnarray}
&&{\cal L}=\partial_\mu\theta \partial^\mu\theta + 
{1\over 1 + 4\cot^2\theta}\biggl({9\over4}\> \partial_\mu\widetilde{\psi}\partial^\mu\widetilde{\psi} +\cot^2\theta\> \partial_\mu\varphi\partial^\mu\varphi
\nonumber\\[5pt]
&&
\hspace{5.5cm}
-{3\over2} \epsilon^{\mu\nu}\> \partial_\mu\widetilde{\psi} \partial_\nu \varphi\biggr)
+ 2\mu_+\mu_- \cos\theta \cos\varphi.
\label{LagrangianEHDual}
\end{eqnarray}
To our knowledge, the relationship between this Lagrangian and the Pohlmeyer reduced ${\fl C}P^2$ nonlinear sigma model or, equivalently, bosonic string theory on ${\fl R}_t\times {\fl C}P^2$ has not been pointed out before. 

The Lagrangians~(\ref{LagrangianEH}) and~(\ref{LagrangianEHDual}) specify different nonlinear sigma models with a potential term. Notice that the target-space metric corresponding to~(\ref{LagrangianEH}) is singular at $\theta=0$. In contrast, the metric corresponding to~(\ref{LagrangianEHDual}) is free of singularities, although it becomes non-invertible at $\theta=0$. In addition, the Lagrangian~(\ref{LagrangianEHDual}) exhibits a non-trivial antisymmetric tensor term that is absent in~(\ref{LagrangianEH}).
Nevertheless, their equations of motion correspond to the same set of SSSG equations, and it is not difficult to check that
they are related by a target-space duality transformation 
like those discussed in~\cite{HSGphases}. It corresponds to the canonical transformation
\begin{equation}
\Pi=-{3\over 2} \partial_x\widetilde{\psi} \quad{\rm and}\quad
\widetilde{\Pi}=-{3\over 2}  \partial_x\psi,
\end{equation}
where $\Pi$ and $\widetilde{\Pi}$ are the canonical momenta associated to $\psi$ and $\widetilde{\psi}$ in~(\ref{LagrangianEH}) and~(\ref{LagrangianEHDual}), respectively. On-shell, the equations of the canonical transformations can also be obtained from the condition that $b_\pm$ and $\theta$ are the same in~(\ref{Newfield1}) and~(\ref{Newfield2}).

The emergence of two different Lagrangians exhibits the non-uniqueness of the Lagrangian formulation of the SSSG equations pointed out in Section~\ref{LAGSSSG}. The construction presented there provides a Lagrangian action for the SSSG equations corresponding to ${\fl C}P^2$ in terms of a $G/H=U(2)/U(1)$ gWZW action with a potential term.
That construction requires to reduce the $U(1)_L\times U(1)_R$ gauge symmetry~(\ref{CPgauge}) using the gauge conditions~(\ref{EoMgWZW2}), which depend on the choice of the two homomorphisms $\epsilon_{L/R}:U(1)\rightarrow U(2)$ constrained by the `anomaly free' condition~(\ref{AnomalyFree}).
In this case, there are only two non-equivalent possibilities. The first one is 
\begin{equation}
\epsilon_L=\epsilon_R: {\rm e}^{i\phi} \rightarrow {\rm e}^{i\phi\> T}
\end{equation}
where $T$ is given by~(\ref{CP2algebra}). The corresponding gauge conditions~(\ref{EoMgWZW2}) are
\begin{eqnarray}
&&
(1+3\cos^2\theta)(a_+ + \partial_+\alpha)+ (1+\cos(2\theta))\partial_+\varphi - 4\partial_+\beta = 4a_+
\nonumber\\[5pt]
&&
{\rm and}\quad
(1+3\cos^2\theta)(a_- + \partial_-\beta)- (1+\cos(2\theta))\partial_-\varphi - 4\partial_+\alpha = 4a_-,
\end{eqnarray}
which can be solved as
\begin{equation}
b_\pm=\pm{1\over 3\sin^2\theta}\left( 2\partial_+(\alpha-\beta) + \cos^2\theta\> \partial_+\varphi\right).
\end{equation}
The comparison with~(\ref{Newfield1}) shows that $\psi=2(\alpha-\beta)$, and that the residual gauge transformations are indeed of `vector type', which corresponds to $\rho_+=\rho_-$ in~(\ref{CPgauge}). Therefore,~(\ref{LagrangianEH}) is the  local Lagrangian that corresponds to the Lagrangian action
\begin{equation}
S_{\rm gWZW}^{(V)}\bigl[\gamma,{\cal A}_\pm\bigr]-{\mu_+\mu_-\over 2\pi} \int d^2 x\> \mathop{\rm Tr}\bigl(\Lambda_+ \gamma^{-1}\Lambda_- \gamma\bigr),
\label{CPActionV}
\end{equation}
where $S_{\rm gWZW}^{(V)}$ denotes the  $U(2)/U(1)$ gWZW action of vector type.

The second choice of the two homomorphisms is 
\begin{equation}
\epsilon_L=\epsilon_R^{-1}: {\rm e}^{i\phi} \rightarrow {\rm e}^{i\phi\> T}
\end{equation}
Then, the gauge conditions~(\ref{EoMgWZW2})  read
\begin{eqnarray}
&&
(1+3\cos^2\theta)(a_+ + \partial_+\alpha)+ (1+\cos(2\theta))\partial_+\varphi - 4\partial_+\beta = -4a_+
\nonumber\\[5pt]
&&
{\rm and}\quad
(1+3\cos^2\theta)(a_- + \partial_-\beta)- (1+\cos(2\theta))\partial_-\varphi - 4\partial_+\alpha = -4a_-,
\end{eqnarray}
and they lead to
\begin{equation}
b_\pm={1\over 1+3\cos^2\theta}\left( 2\partial_+(\alpha+\beta) \mp \cos^2\theta\> \partial_+\varphi\right).
\end{equation}
Looking now at~(\ref{Newfield2}), it shows that $\widetilde\psi=2(\alpha+\beta)$, and that the residual gauge transformations are of `axial type'; namely, they corresponds to $\rho_+=-\rho_-$ in~(\ref{CPgauge}). Therefore, the Lagrangian~(\ref{LagrangianEHDual}) corresponds to the Lagrangian action
\begin{equation}
S_{\rm gWZW}^{(A)}\bigl[\gamma,{\cal A}_\pm\bigr]-{\mu_+\mu_-\over 2\pi} \int d^2 x\> \mathop{\rm Tr}\bigl(\Lambda_+ \gamma^{-1}\Lambda_- \gamma\bigr),
\label{CPActionA}
\end{equation}
where $S_{\rm gWZW}^{(A)}$ denotes now the  $U(2)/U(1)$ gWZW action of axial type.
The two actions~(\ref{CPActionV}) and~(\ref{CPActionA})
have a global $U(1)$ symmetry that gives rise to a target-space duality symmetry that relates them off-shell~\cite{HSGphases}. It coincides with the transformation that relates the two Lagrangians~(\ref{LagrangianEH}) and~(\ref{LagrangianEHDual}).

\subsection{Pohlmeyer reduction of principal chiral models: The HSG theories}
\label{ExampleChiralModel}

We shall now illustrate the variety of SSSG equations associated to a single symmetric space of rank larger than~1 by discussing 
the Pohlmeyer reduction of nonlinear sigma models with target-space a symmetric space of type~II; namely, $F/G=G\times G/G_{\rm D}$ with $G$ is a compact simple Lie group. In the following, and just for simplicity, we will also assume that $G$ is simply laced.
In this case, $G\times G=\{(g_1,g_2); \; g_1,g_2\in G\}$ and $G_{\rm D}=\{(\gamma,\gamma); \; \gamma\in G\}$, which exhibits that $G_{\rm D}$ is trivially isomorphic to $G$. Moreover, the action of the gauge transformations~(\ref{Hgauge}) on $G\times G$ is $(g_1,g_2)\rightarrow (g_1\gamma,g_2\gamma)$, and it is straightforward to check that $G\times G/G_{\rm D}$ is also isomorphic to $G$, with the isomorphism given by the map $(g_1,g_2) \rightarrow g_1g_2^{-1}$. Therefore, $G\times G/G_{\rm D}$ is isomorphic to $G$, and the sigma model with target-space $G\times G/G_D$ is just the principal chiral model corresponding to $G$.

The Lie algebra of $F=G\times G$ is ${\go f}={\go g}\oplus {\go g}$, where ${\go g}$ is the Lie algebra of $G$. Its elements are pairs $(a,b)$ with $a,b\in {\go g}$, and the decomposition~(\ref{SSalgebra}) takes the form
\begin{equation}
{\go g}_{\rm D}=\{ (a,a)\>:\> a\in {\go g}\}
\quad{\rm and}\quad
{\go p}=\{ (a,-a)\>:\> a\in {\go g}\},
\end{equation}
which is orthogonal with respect to $\mathop{\rm Tr}\bigl((a,b)\cdot(c,d)\bigr)=\mathop{\rm Tr}(ac)+\mathop{\rm Tr}(bd)$. 
As explained in Section~(\ref{SSSG}), Pohlmeyer reduction involves the choice of two constant elements $\Lambda_+$ and $\Lambda_-$ in a fixed maximal abelian subspace ${\go a}\subset {\go p}$. In this case, it is not difficult to show that the maximal abelian subspaces of ${\go p}$ are in one-to-one correspondence with the Cartan subalgebras of ${\go g}$. Namely, if ${\go s}\subset {\go g}$  is a Cartan subalgebra, the corresponding maximal abelian subspace is
\begin{equation}
{\go a}= \{(a,-a)\>:\> a\in {\go s}\}\subset{\go p},
\end{equation}
which exhibits that $\mathop{\rm rank}(G\times G/G_{\rm D})=\mathop{\rm rank}(G)$.

Let us introduce a Cartan-Weyl basis for the complexification of ${\go g}$, which consists of a fixed Cartan subalgebra ${\go s}$ with $r=\mathop{\rm rank}(G)$ generators
$\{h^1,\ldots, h^r\}$ and step operators $E_{\vec{\alpha}}$ for each root ${\vec{\alpha}}$, so that they satisfy
\begin{equation}
\bigl[\>\vec{\lambda}\cdot\vec{h},\> E_{\vec{\alpha}}\>\bigr]\> =\> (\vec{\lambda}\cdot\vec{\alpha})\> E_{\vec{\alpha}} \quad{\rm and} \quad
\bigl[\> E_{\vec{\alpha}},\> E_{-\vec{\alpha}}\>\bigr]\> =\>  \vec{\alpha}\cdot \vec{h}.
\end{equation}
In terms of this basis, ${\go g}$ is spanned by the (anti-Hermitian) generators $ih^1,\ldots, ih^r$, $i(E_{\vec{\alpha}}+ E_{-\vec{\alpha}})$ and $E_{\vec{\alpha}}- E_{-\vec{\alpha}}$. Then, the two constant elements of ${\go a}$ required to perform the reduction can be written as
\begin{equation}
\Lambda_+=(\vec{\lambda}_+\cdot \vec{h},\> -\vec{\lambda}_+\cdot \vec{h})\quad {\rm and}\quad
\Lambda_-=(\vec{\lambda}_-\cdot \vec{h},\> -\vec{\lambda}_-\cdot \vec{h}),
\end{equation}
where $\vec{\lambda}_+$ and $\vec{\lambda}_-$ are two real $r$-dimensional vectors.
In the following, it will be convenient to fix a basis of simple roots $\{\vec{\alpha}_1,\ldots,\vec{\alpha}_r\}$ and the corresponding basis of fundamental weights $\{\vec{\lambda}_1,\ldots,\vec{\lambda}_r\}$, so that
$\vec{\lambda}_i\cdot\vec{\alpha}_j=\delta_{ij}$. Then, the two vectors can be written as
\begin{equation}
\vec{\lambda}_+ = \sum_{i=1}^r m_i^+ \vec{\lambda}_i
\quad{\rm and}\quad
\vec{\lambda}_- = \sum_{i=1}^r m_i^- \vec{\lambda}_i,
\label{LambdasWW}
\end{equation}
where $m_1^+,\ldots,m_r^+$ and $m_1^-,\ldots,m_r^-$ are real constants. In this case, the reduction procedure described in Section~\ref{SSSG} gives rise to a different set of SSSG equations for each non-equivalent choice of those constant coefficients. We will consider just three choices that lead to rather different sets of equations.

The first one corresponds to~\footnote{This case has been recently considered in~\cite{GTseytlin2}.}
\begin{equation}
m_i^+,m_i^- \not=0 \quad \forall\; i=1,\ldots,r,
\label{NonDegenerate}
\end{equation}
which ensures that $\vec{\lambda}_+$ and $\vec{\lambda}_-$ are not orthogonal to any root of ${\go g}$.
Then, $H^{(+)}=H^{(-)}$ is the maximal torus $U(1)^r$ of $G_{\rm D}\simeq G$ associated to the Cartan subalgebra ${\go s}$.
The resulting SSSG equations are the equations of motion of the so-called homogeneous sine-Gordon (HSG) theories~\cite{HSG}, which are two-dimensional integrable theories whose classical and quantum properties have been extensively studied in the literature~\cite{HSGplus1,HSGplus2,HSGplus3}. Their Lagrangian formulation is provided by a gWZW action corresponding to the coset $G/U(1)^r$ with a potential term fixed by $\Lambda_+$ and $\Lambda_-$.

This first case exhibits two interesting features of the reduced models associated to symmetric spaces of rank larger than~1. The first one is that the resulting set of integrable equations depends on adjustable parameters that play the role of coupling constants. This is so because the form of $H^{(+)}$ and $H^{(-)}$ is independent of the precise value of the parameters $m_i^+$ and $m_i^-$ in~(\ref{LambdasWW}) as far as the conditions~(\ref{NonDegenerate}) are satisfied. Those parameters determine the mass spectrum of the theory. The spectrum of fundamental particles can be easily obtained by studying the linearized form of the gauge-fixed equations~(\ref{ENAAT}) around the obvious vacuum configuration $\gamma_0=1$; namely,
\begin{equation}
\gamma={\rm e}^{\> i \phi}\approx 1+ i\phi\;\Rightarrow
\begin{cases}
\;\partial_+\partial_-\phi= \mu^2\> \bigl[\Lambda_+,[\Lambda_-, \phi]\bigl],
& \\[5pt]
\;{\rm and}\quad {\mathop{\rm P}}_{{\go h}_+}\bigl(\partial_+\phi\bigr) ={\mathop{\rm P}}_{{\go h}_-}\bigl( \partial_-\phi\bigr) = 0,&
\end{cases}
\label{Fluctua}
\end{equation}
where $\phi\in {\go g}$. Then, for each root $\vec{\alpha}$, the configuration $\phi=\psi E_{\vec{\alpha}} -\psi^\ast E_{-\vec{\alpha}}$ corresponds to a fundamental particle of mass~\footnote{
Since $m_{\vec{\alpha}}^2$ has to be positive, $\gamma_0=1$ corresponds to a true vacuum configuration only if all the constant parameters
$m_1^\pm,\ldots,m_r^\pm$ are of the same sign~\cite{HSGphases,HSGVall}.}
\begin{equation}
m_{\vec{\alpha}}^2= \mu^2 (\vec{\alpha}\cdot \vec{\lambda}_+)(\vec{\alpha}\cdot \vec{\lambda}_-)\not=0,
\end{equation}
whose value depends on $m_i^+$ and $m_i^-$. This particle carries a $U(1)^r$ Noether charge whose value is characterised by ${\vec{\alpha}}$. The equations of motion of the HSG theories also admit soliton solutions whose masses are also determined by the parameters $m_i^\pm$~\cite{HSGsol}.

The second feature is the non-uniqueness of the Lagrangian formulation, which involves the choice of the homomorphisms $\epsilon_L$ and $\epsilon_R$ in~(\ref{Embed}). For the HSG theories, it is customary to choose them such that $\epsilon_L=1$ and $\epsilon_R=\widehat\tau$~\cite{HSG}, where $\widehat\tau$ can be any element of $\Lambda^\ast_w(G)$, the (discrete) group of automorphism of the dual lattice to the weight lattice of $G$~\cite{HSGphases}.
The relationship between the different $\widehat\tau$-dependent Lagrangian formulations has been recently investigated making use of target-space duality symmetries, with the result that not all the Lagrangian theories seem to be related by standard T-duality transformations for $G=SU(n)$ with $n\geq5$ and $E_6$~\cite{HSGVall}. It would be interesting to revise those results on the light of the interpretation of~(\ref{EoMgWZW2}) as gauge conditions.

Other non-equivalent possible choices of the parameters in~(\ref{LambdasWW}) are obtained by making some of them vanish. Since our purpose here is just to illustrate the differences between the resulting sets of equations, we will now restrict ourselves to $G=SU(3)$ and use its defining representation in terms of $3\times 3$ unitary matrices. Then, we will first consider the choice $m_i^+=m_i^- =\delta_{i,2}$ in~(\ref{LambdasWW}), which is the limit $m_1^+, m_1^-\rightarrow 0$ of~(\ref{NonDegenerate}). It corresponds to
\begin{equation}
\Lambda_+=\Lambda_-={1\over3}
\left(\begin{array}{ccc}1 & 0 & 0 \\0 & 1 & 0 \\0 & 0 & -2\end{array}\right),
\end{equation}
whose centralisers in $SU(3)$ are of the form
\begin{equation}
H^{(+)}=H^{(-)}=
\left(\begin{array}{ccc}\ast & \ast& 0 \\ \ast& \ast & 0 \\0 & 0 & \ast\end{array}\right)\simeq U(2)\>.
\end{equation}
Therefore, the Lagrangian formulation of the resulting SSSG equations is given by a gWZW action corresponding to the coset $SU(3)/U(2)$ with a potential term. The spectrum of fundamental particles can be obtained by studying the linearized equations~(\ref{Fluctua}), which show that the configurations of the form
\begin{equation}
\phi=
\left(\begin{array}{ccc}0 & 0& \phi_{13}(\tau,x) \\ 0& 0 & \phi_{23}(\tau,x) \\-\phi_{13}^\ast(\tau,x) & -\phi_{23}^\ast(\tau,x) &0\end{array}\right),
\end{equation}
describe a $U(2)$ multiplet of particles of equal mass fixed by $\mu$.

Another interesting non-equivalent choice of the parameters is $m_i^+=\delta_{i,2}$ and 
$m_i^- =\delta_{i,1}$, which is the limit $m_1^+,m_2^-\rightarrow0$ of~(\ref{NonDegenerate}). It corresponds to
\begin{equation}
\Lambda_+={1\over3}
\left(\begin{array}{ccc}1 & 0 & 0 \\0 & 1 & 0 \\0 & 0 & -2\end{array}\right)
\quad {\rm and}\quad
\Lambda_-={1\over3}
\left(\begin{array}{ccc}2 & 0 & 0 \\0 & -1 & 0 \\0 & 0 & -1\end{array}\right)\>,
\end{equation}
whose centralisers are now of the form
\begin{equation}
H^{(+)}=
\left(\begin{array}{ccc}\ast & \ast& 0 \\ \ast& \ast & 0 \\0 & 0 & \ast\end{array}\right)
\quad {\rm and}\quad
H^{(-)}=
\left(\begin{array}{ccc}\ast & 0& 0 \\ 0& \ast & \ast \\0 & \ast & \ast\end{array}\right)\>.
\end{equation}
Therefore, in this case $H^{(+)}\not=H^{(-)}$, but both $H^{(+)}$ and $H^{(-)}$ are isomorphic to $U(2)$. 
Then, the Lagrangian formulation of the corresponding SSSG equations is also provided by a gWZW action associated to the coset $SU(3)/U(2)$ with a potential term, but now $\epsilon_L$ is necessarily $\not=\epsilon_R$ and the relevant coset is always asymmetric. Clearly, this case falls outside the class considered by Bakas, Park and Shin in~\cite{BPS}.
Once more, the spectrum of fundamental particles can be obtained by studying the linearized equations~(\ref{Fluctua}). In this case,
\begin{equation}
\phi=
\left(\begin{array}{ccc}0 & \phi_{12}(x_-)& \phi_{13}(\tau,x) \\ -\phi_{12}^\ast(x_-)& 0 & \phi_{23}(x_+) \\-\phi_{13}^\ast(\tau,x) & -\phi_{23}^\ast(x_+) &0\end{array}\right)
\end{equation}
describes a massive particle of mass $\mu$ associated to $\phi_{13}$, and two massless particles associated to $\phi_{12}$ (right-mover) and $\phi_{23}$ (left-mover). 

\section{SSSG equations from sigma models with target-space \boldmath{$AdS_n$}}
\label{AdS}

When the symmetric space $F/G$ is of indefinite signature, $G$ is noncompact and the polar coordinate decomposition used in Section~\ref{SSSG} to solve the constraints~(\ref{GenConstraint}) is not satisfied anymore. In fact, the general theory of symmetric spaces is very extensive~\cite{Neill,Figueroa} and we are not aware of any general procedure to solve them. Here, we will just consider the SSSG equations associated to the anti-de~Sitter spaces $AdS_n$, which exhibit some features different to those of the equations
corresponding to symmetric spaces of definite signature discussed in Section~\ref{SSSG}.
One of them concerns the identification of the conditions~(\ref{StressTensor}) with the Virasoro constraints of a classical bosonic string theory. When ${\cal M}=F/G$ is of definite signature, $T_{++}$ and $T_{--}$ are positive definite and, consequently, $\mu^2>0$ in~(\ref{StressTensor}). Then, the relevant curved space-time is always ${\fl R}_t\times {\cal M}$~\cite{Tseytlin2003}. In contrast, 
if the signature of $F/G$ is indefinite, the sign of $T_{++}$ and $T_{--}$ is not definite and $\mu^2$ is not constrained to be positive anymore. This enables the construction of 
SSSG equations corresponding to
$\mu^2<0$ and $\mu^2=0$ that can be used to describe bosonic string configurations on ${\cal M}\times S^1_\vartheta$~\cite{GTseytlin1} and on ${\cal M}$~\cite{DeVega,Jevicki}, respectively. Obviously, the case $\mu^2=0$ should be expected to be different to the others, since the corresponding constraints do not break the (classical) conformal invariance of the sigma model. In fact, only the reductions with $\mu^2\not=0$ follow the pattern summarised by 
Figure~\ref{Figura}. 
In particular, the SSSG equations corresponding to $\mu^2=0$ are not of NAAT type, and the construction of Section~\ref{LAGSSSG} cannot be used to find their Lagrangian formulation.

The anti-de~Sitter space
\begin{eqnarray}
AdS_n&=& \bigl\{(x_1,\ldots,x_{n+1})\in {\fl R}^{n+1}\>:\> -x_1^2 -x_2^2 + x_3^2+\cdots +x_{n+1}^2=-1\bigr\}\nonumber\\[5pt]
&=&
SO(2,n-1)/SO(1,n-1)
\label{AdSdefinition}
\end{eqnarray}
is a symmetric space of Lorentzian $(1,n-1)$ signature. 
It is worth noticing that $SO(2,n-1)$ is non-connected (it has two different components), and that only the identity component has to be considered in~(\ref{AdSdefinition})~\cite[Chapter~11]{Neill}. 
Using the $(n+1)\times (n+1)$ matrix representation of $SO(2,n-1)$ and its diagonally embedded $SO(1,n-1)$ subgroup, the form of the elements $r\in{\go g}$ and $k\in{\go p}$ in~(\ref{SSalgebra}) is
\begin{equation}
r=
\left(\begin{array}{ccccc}
0 & 0 & 0 & \cdots & 0 \\
0 & 0 & a_1 & \cdots & a_{n-1} \\
0 & a_1 &  &  &  \\
\vdots & \vdots &  & {\cal N} &  
\\0 & a_{n-1} &  &  & 
\end{array}
\right)
\quad {\rm and}\quad
k=
\left(\begin{array}{ccccc}
0 &- v_0 & v_1 & \cdots & v_{n-1} \\
v_0 & 0 & 0 & \cdots & 0 \\
v_1 & 0 & 0 & \cdots & 0\\
\vdots & \vdots &\vdots  &  & \vdots \\
v_{n-1} & 0 & 0 &\cdots  & 0
\end{array}
\right)\equiv \hat{k}[\bfm{v}],
\end{equation}
where~${\cal N}=-{\cal N}^{\rm T}$ denotes a $(n-1)\times (n-1)$ skew-symmetric matrix and $\bfm{v}=(v_0,v_1,\ldots,v_{n-1})^{\rm T}$ is a real $n$-dimensional column vector. It is straightforward to check that this decomposition is orthogonal with respect to the trace form and, moreover, that the rank of $AdS_n$ is~1. The Lagrangian of the nonlinear sigma model with target-space $AdS_n$ is of the form~(\ref{Lagrangian}) with $\kappa=-1$; namely,
\begin{equation}
{\cal L}= +{1\over2} \mathop{\rm Tr} \bigl(J_\mu J^\mu\bigr)\;
\Rightarrow\; T_{\pm\pm}= +{1\over2} \mathop{\rm Tr} \bigl(J_\pm J_\pm\bigr),
\label{LagrangianAdS}
\end{equation}
so that the contribution of the spacelike configurations to $T_{\pm\pm}$ is positive definite.

In order to solve the constraints~(\ref{GenConstraint}), we shall proof an analogue of the polar coordinate decomposition satisfied by the symmetric spaces of definite signature. It is motivated by the explicit proof of the polar coordinate decomposition for $S^n$ presented at the beginning of Appendix~\ref{AppCSG}. Consider a generic element of $SO(1,n-1)\subset SO(2,n-1)$~\footnote{The diagonally embedded $SO(1,n-1)$ subgroup of $SO(2,n-1)$ is the isotropy group of the point $(1,0,\ldots,0)\in AdS_n$.},
\begin{equation}
g=\left(\begin{array}{cc}
1 & 0\;\\
0& \;N^{-1}\end{array}\right)
\quad {\rm with}\quad
N\in SO(1,n-1)\>.
\label{AdSreduced}
\end{equation}
The transformation $\hat{k}[\bfm{v}]\rightarrow g^{-1} \hat{k}[\bfm{v}] g$ amounts to $\bfm{v}\rightarrow N\bfm{v}$, which is a $(1,n-1)$ dimensional Lorentz transformation of the vector $\bfm{v}$ that, by definition, preserves the quadratic form
$\mathop{\rm Tr}(\hat{k}^2[\bfm{v}])/2=-v_0^2 + v_1^2 +\cdots+ v_{n-1}^2$. This provides a natural classification of the elements $k\in{\go p}$ according to the sign of $\mathop{\rm Tr}(k^2)$: 
`spacelike' if $\mathop{\rm Tr}(k^2)>0$, `timelike' if $\mathop{\rm Tr}(k^2)<0$, and `lightlike' if $\mathop{\rm Tr}(k^2)=0$. Then, using well known properties of the Lorentz group, in each case it is possible to construct a transformation that takes the file vector to some specific canonical form. Namely,
\begin{eqnarray}
&&
(v_0,v_1,\dots,v_{n-1}) \rightarrow  \mu(0,0,0,\ldots,1)\quad {\rm if}\quad \mathop{\rm Tr}(\hat{k}^2[\bfm{v}])=+2\mu^2>0\nonumber\\[5pt]
&&
(v_0,v_1,\dots,v_{n-1})  \rightarrow \mu(1,0,0,\ldots,0)\quad {\rm if}\quad \mathop{\rm Tr}(\hat{k}^2[\bfm{v}])=-2\mu^2<0\nonumber\\[5pt]
&&
(v_0,v_1,\dots,v_{n-1}) \rightarrow \mu(1,0,\ldots,0,1)\quad {\rm if}\quad \mathop{\rm Tr}(\hat{k}^2[\bfm{v}])=0,
\label{LorentzTransf}
\end{eqnarray}
where $\mu$ denotes a real number. In addition, for $AdS_2$ there is a fourth possibility:
\begin{equation}
(v_0,v_1) \rightarrow \mu(1,-1)\quad {\rm if}\quad \mathop{\rm Tr}(\hat{k}^2[\bfm{v}])=0,
\end{equation}
which is non-equivalent to the others because
$(1,-1)^{\rm T}\not=N(1,1)^{\rm T}$ for any $N\in SO(1,1)$.

Taking all this into account, the analogue of the polar coordinate decomposition for $AdS_n$ can be stated as follows. For any $k\in{\go p}$ there exists $\overline{g}\in SO(1,n-1)$ and $\mu\in{\fl R}$ such that
\begin{equation}
\overline{g}^{\> -1} k \overline{g}
=\begin{cases}
\mu\> \widehat{k}[(0,\ldots,0,1)]\equiv \mu\> T^{(s)}, & \text{\rm if}\quad \mathop{\rm Tr}(k^2)>0,\\[3pt]
\mu\> \widehat{k}[(1,0,\ldots,0)]\equiv \mu\> T^{(t)}, & \text{\rm if}\quad \mathop{\rm Tr}(k^2)<0, \\[3pt]
\mu\> \widehat{k}[(1,0,\ldots,0,1)]\equiv \mu\> T^{(l)}, & \text{\rm if}\quad \mathop{\rm Tr}(k^2)=0.
\end{cases}
\label{PCDads}
\end{equation}
Furthermore, for $AdS_2$ there is the fourth non-equivalent possibility
\begin{equation}
\overline{g}^{\> -1} k \overline{g}
=\mu\> \widehat{k}[(1,-1)]\equiv \mu\> \widetilde{T}^{(l)}, \quad{\rm if}\quad \mathop{\rm Tr}(k^2)=0.
\label{AdS2plus}
\end{equation}
It is worth comparing~(\ref{PCDads}) with~(\ref{PCDsimple}). Notice that ${\go a}^{(s)}={\fl R}\> T^{(s)}$, ${\go a}^{(t)}={\fl R}\> T^{(t)}$  and ${\go a}^{(l)}={\fl R}\> T^{(l)}$ are three maximal (one-dimensional) abelian subspaces of ${\go p}$ which, according to~(\ref{PCDads}), are not conjugated under the adjoint action of $G=SO(1,n-1)$. This is in contrast to the case of symmetric spaces $F/G$ of definite signature, where the polar coordinate decomposition ensures that all the maximal abelian subspaces ${\go a}\subset {\go p}$ are conjugated under the adjoint action of $G$ (see Section~\ref{SSSG}).
The generalised decomposition~(\ref{PCDads}) exhibits that for $AdS_n$ the constraints $\mathop{\rm Tr}(J_\pm^2)={\rm constant}$ are the only independent ones in~(\ref{GenConstraint}) or, in other words, that the only `primitive' chiral densities are $T_{++}$ and $T_{--}$ akin to the case of sigma models with target-space a symmetric space of definite signature and rank~1~\cite{EvansMountain}. 

The decomposition summarised by~(\ref{PCDads}) makes possible to construct the SSSG equations corresponding to the sigma models with target-space $AdS_n$ by following the procedure of Section~3. 
First, taking~(\ref{EoM2}) into account, we can apply~(\ref{PCDads}) to the currents $J_\pm$ so that the general solution to the constraints~(\ref{GenConstraint}) is of the form
\begin{equation}
J_\pm = \mu_{\pm}\> \overline{g}_{\pm} \Lambda_\pm \overline{g}_\pm^{\>-1},
\label{GenPCD}
\end{equation}
where $\overline{g}_\pm\in SO(1,n-1)$, $\mu_+$ and $\mu_-$ are real numbers, and $\Lambda_+$ and $\Lambda_-$ are constant and equal to either $T^{(s)}$, or $T^{(t)}$, or $T^{(l)}$ (or $\widetilde{T}^{(l)}$ for $AdS_2$). Then,~(\ref{GaugeJ}) and~(\ref{GenPCD}) enable the `partial reduction' gauge condition~(\ref{RedCurrents}),
$$
J_+ = \mu_+ \Lambda_+\quad {\rm and}\quad J_- = \mu_- \gamma^{-1} \Lambda_- \gamma,
$$
where $\gamma=\overline{g}_-^{\>-1}\overline{g}_+\in SO(1,n-1)$.
In the rest of this section, we shall discuss the equations corresponding to $\Lambda_+=\Lambda_-$. We will also work out the explicit form of the SSSG equations corresponding to $AdS_2$ and $AdS_3$.

\subsection{`Spacelike' reduction: Bosonic strings on ${\fl R}_t\times AdS_n$}

We start with
\begin{equation}
\Lambda_+=\Lambda_-=T^{(s)},
\end{equation}
which is the solution to the constraints
\begin{equation}
T_{++} = +\mu_+^2 >0
\quad{\rm and}\quad 
T_{--} = +\mu_-^2 >0.
\label{AdSConstPos}
\end{equation}
Then, provided that $\mu_+^2=\mu_-^2=\mu^2$, the solutions to the corresponding SSSG equations describe bosonic string configurations in ${\fl R}_t\times AdS_n$ using the orthonormal gauge condition $t=\mu\tau$~\cite{Tseytlin2003}. In this case, the elements in the centraliser of $T^{(s)}=\widehat{k}[(0,\ldots,0,1)]$ in $G=SO(1,n-1)$ are of the form
\begin{equation}
\left(\begin{array}{ccc}
1 & 0\;&0\;\\
0&\;N&0\\
0& 0&1\end{array}\right),
\quad {\rm with}\quad N\in SO(1,n-2)\>
\end{equation}
which shows that $H^{(+)}=H^{(-)}$, and that both are isomorphic to $SO(1,n-2)$. Then, following~\cite{BPS} and Section~\ref{LAGSSSG}, the Lagrangian formulation of the resulting SSSG equations is provided by a gWZW action associated to the coset $SO(1,n-1)/SO(1,n-2)$ with a potential term. In this case the reduction follows the pattern summarised by figure~\ref{Figura}.

The simplest example corresponds to $n=2$, where 
\begin{equation}
\Lambda_+=\Lambda_-=\widehat{k}[(0,1)]=\left(\begin{array}{ccc}
0 & 0&1\;\\
0&0&0\\
1& 0&0
\end{array}\right).
\end{equation}
Since the field $\gamma$ takes values in (the identity component of) $SO(1,1)$, which is a one-parameter (abelian) group, it can be parameterised as
\begin{equation}
\gamma=\left(\begin{array}{ccc}
1 & 0&0\;\\
0&\cosh\chi&\sinh\chi\\
0& \sinh\chi&\cosh\chi
\end{array}\right) = \exp \left(\begin{array}{ccc}
0& 0&0\;\\
0&0&\chi\\
0& \chi&0
\end{array}\right)
\label{AdS2}
\end{equation}
in terms of a real field~$\chi$, and it is straightforward to check that 
$H^{(+)}$ and $H^{(-)}$ are trivial. The resulting SSSG equation is the well-known sinh-Gordon equation
\begin{equation}
\partial_+\partial_-\chi -\mu_+\mu_-\sinh\chi=0,
\end{equation}
which is the equation of motion of the Lagrangian
\begin{equation}
{\cal L}=\partial_+\chi \partial_-\chi +\mu_+\mu_-\cosh\chi.
\end{equation}
It is worth noticing that the constraints~(\ref{AdSConstPos}) do not fix the sign of $\mu_+$ and $\mu_-$ and, in fact,  
the potential is bounded from below only if we take $\mu_+\mu_-<0$.

The first case with non-trivial gauge groups $H^{(+)}$ and $H^{(-)}$ is provided by the reduction of the $AdS_3$ sigma model. Then
\begin{equation}
\Lambda_+=\Lambda_-=\widehat{k}[(0,0,1)]=\left(
\begin{array}{cccc}
0&0&0&1\\
0&0&0&0\\
0&0&0&0\\
1&0&0&0
\end{array}\right)
\end{equation} 
and, using the notation
\begin{equation}
b_1=\left(
\begin{array}{cccc}
0&0& 0 & 0 \\
0&0&1&0\\
0&1&0&0\\
0&0&0&0
\end{array}\right)=+b_1^{\rm T}
\quad {\rm and}\quad
r=\left(
\begin{array}{cccc}
0&0& 0 & 0 \\
0&0&0&0\\
0&0&0&1\\
0&0&-1&0
\end{array}\right)=-r^{\rm T\>},
\label{BoostRot}
\end{equation}
the centraliser of $\Lambda_+=\Lambda_-$ in $SO(1,2)$ is
\begin{equation}
H^{(+)}=H^{(-)}=\{{\rm e}^{\>x\> b_1}; \;x\in {\fl R} \} 
\simeq SO(1,1).
\end{equation}
This motivates the parameterisation
\begin{equation}
\gamma = {\rm e}^{\>\alpha\> b_1}\> {\rm e}^{\>\theta\> r}\> {\rm e}^{\>-\beta\> b_1} \in SO(1,2),
\end{equation}
in terms of three real fields $\alpha$, $\beta$ and $\theta$. Correspondingly, the gauge fields in~(\ref{RedGauge}) are
\begin{equation}
A_-^{(R)} = a_-\> b_1\quad {\rm and}\quad A_+^{(L)} = a_+\> b_1,
\end{equation}
with $a_\pm \in{\fl R}$, and the $SO(1,1)_L\times SO(1,1)_R$ gauge transformations~(\ref{ESSSG1}) read
\begin{equation}
\alpha\rightarrow \alpha + \rho_-, \quad
\beta\rightarrow \beta + \rho_+, \quad{\rm and}\quad
a_\pm \rightarrow a_\pm -\partial_\pm \rho_{\mp},
\label{GaugeAdSspace}
\end{equation}
where $h_\pm =  {\rm e}^{\>\rho_\pm b_1}$.
Then, in terms of the three gauge invariant fields 
\begin{equation}
\theta,\quad b_+= a_+ +\partial_+\alpha\quad {\rm and}\quad b_-= a_- +\partial_-\beta,
\end{equation}
the zero-curvature equations of motion~(\ref{ESSSG2}) become
\begin{eqnarray}
&&\partial_+\partial_-\theta + (b_+b_- - \mu_+\mu_-)\sin\theta=0
\label{AdSspaceone}\\[5pt]
&&\partial_+\bigl((1+\cos\theta)b_-\bigr) -\partial_-\bigl((1+\cos\theta)b_+\bigr)=0
\label{AdSspacetwo}\\[5pt]
&&\partial_+\bigl((1-\cos\theta)b_-\bigr) +\partial_-\bigl((1-\cos\theta)b_+\bigr)=0.
\label{AdSspacethree}
\end{eqnarray}
They are related to the SSSG equations~(\ref{Sone}--\ref{Sthree}) corresponding to $SO(3)$ by means of the analytic continuation $b_\pm \rightarrow ib_\pm$.
According to Section~\ref{LAGSSSG}, these SSSG equations admit a Lagrangian formulation in terms of a $SO(1,2)/SO(1,1)$ gWZW action with a potential term. 
Since $SO(1,1)$ is a one-parameter (abelian) group, like $U(1)$ in~(\ref{CP2algebra}) or $SO(2)$ in~(\ref{SO2form}),
there are two different Lagrangian actions, of `axial' or `vector' type, related by a target-space duality symmetry 
like those discussed in~\cite{HSGphases}. They give rise to two local Lagrangians that can obtained directly from the SSSG equations. They are of the general form
\begin{equation}
{\cal L}\bigl[U,V,\lambda\bigr]= {\partial_\mu
U\>\partial^\mu V\over 1- U V} - \lambda\>
U V
\label{AdSspaceLag}
\end{equation}
where $U$ and $V$ are complex fields subjected to specific reality conditions. Notice that this Lagrangian specifies different analytic continuations of the complex sine-Gordon Lagrangian~(\ref{CSGLag}), which corresponds to $U=+V^\ast$. 

The procedure to establish the correspondence between the SSSG equations~(\ref{AdSspaceone}--\ref{AdSspacethree}) and the equations of motion of~(\ref{AdSspaceLag}) is identical to the one used to find the Lagrangians of the reduced ${\fl C}P^2$ and $S^3$ sigma models in Section~\ref{ExampleCPn} and~Appendix~\ref{AppCSG}, respectively.
First, we use~(\ref{AdSspacetwo}) to write $b_+$ and $b_-$ in terms of a new real field $\varrho$,
\begin{equation}
b_\pm = {2\over 1+\cos\theta}\> \partial_\pm \varrho.
\end{equation}
Then,~(\ref{AdSspaceone}) and~(\ref{AdSspacethree}) become
\begin{eqnarray}
&&
\partial_\mu \bigl(\tan^2(\theta/2) \partial^\mu\varrho\bigr)=0\nonumber\\[5pt]
&&
\partial_+\partial_-\theta +{4\sin\theta\over(1+\cos\theta)^2}\> \partial_+\varrho\partial_-\varrho +\mu_+\mu_-\sin\theta=0,
\end{eqnarray}
which are the equations of motion of~(\ref{AdSspaceLag}) with
\begin{equation}
U=\sin(\theta/2) {\rm e}^{\> \varrho}, \quad
V=\sin(\theta/2) {\rm e}^{\> -\varrho}
\quad {\rm and} \quad \lambda=+\mu_+\mu_-;
\label{AdsSpaceLagD1}
\end{equation}
namely,
\begin{equation} 
{\cal L}={1\over4} \partial_\mu\theta \partial^\mu\theta - \tan^2(\theta/2)\partial_\mu\rho \partial^\mu\rho -\mu_+\mu_- \sin^2(\theta/2).
\end{equation}
In a completely equivalent way, we can use~(\ref{AdSspacethree}) to write
\begin{equation}
b_\pm =\pm {2\over 1-\cos\theta}\> \partial_\pm \widetilde\varrho,
\end{equation}
which also leads to the equations of motion of~(\ref{AdSspaceLag}) but, in this second case, 
\begin{equation}
U=\cos(\theta/2) {\rm e}^{\> \widetilde\varrho}, \quad
V=\cos(\theta/2) {\rm e}^{\> -\widetilde\varrho}
\quad {\rm and} \quad \lambda=-\mu_+\mu_-,
\label{AdsSpaceLagD2}
\end{equation}
that corresponds to
\begin{equation} 
{\cal L}={1\over4} \partial_\mu\theta \partial^\mu\theta - \cot^2(\theta/2)\partial_\mu\widetilde{\rho} \partial^\mu\widetilde{\rho} +\mu_+\mu_- \cos^2(\theta/2).
\end{equation}
The parameterisations~(\ref{AdsSpaceLagD1}) and~(\ref{AdsSpaceLagD2}) are solutions to the reality conditions
\begin{equation}
U^\ast = U, \quad V^\ast = V \quad{\rm and}\quad 0\leq UV\leq1,
\label{RealitySpace}
\end{equation}
so that the Lagrangian~(\ref{AdSspaceLag}) subjected to them is related to the complex sine-Gordon Lagrangian~(\ref{CSGLag})
by means of the analytic continuation
\begin{equation}
\psi \rightarrow U \in {\fl R}\quad {\rm and}\quad \psi^\ast \rightarrow V\in {\fl R}
\end{equation}
or, equivalently, $\phi\rightarrow i\varrho$ and $\widetilde{\phi}\rightarrow i\widetilde{\varrho}$ in~(\ref{CSGLagD1}) and~(\ref{CSGLagD2}), respectively. It is a generalization of the sine-Gordon Lagrangian with an  internal dilatation symmetry $U\rightarrow \lambda U$, $V\rightarrow \lambda^{-1} V$, $\lambda\in{\fl R}$.
 
\subsection{`Timelike' reduction: Bosonic strings on $AdS_n\times S_\vartheta^1$}
\label{Timelike}

Next, we shall consider
\begin{equation}
\Lambda_+=\Lambda_-=T^{(t)},
\end{equation}
which is the solution to the constraints
\begin{equation}
T_{++} = -\mu_+^2 <0
\quad{\rm and}\quad 
T_{--} = -\mu_-^2 <0.
\end{equation}
Provided that $\mu_+^2=\mu_-^2=\mu^2$, the solutions to the corresponding SSSG equations describe bosonic string configurations on the curved space-time $AdS_n\times S_\vartheta^1$ using the gauge condition $\vartheta=\mu\tau$, where $\vartheta$ is the $S^1$ angular coordinate.
This type of reduction of $AdS_n$ sigma models is the relevant one in the generalisation of Pohlmeyer reduction proposed in~\cite{GTseytlin1,GTseytlin2,MikhailovSS}.
The elements in the centraliser of $T^{(t)}=\widehat{k}[(1,0,\ldots,0)]$ in $G=SO(1,n-1)$ are now of the form
\begin{equation}
\left(\begin{array}{ccc}
1 & 0&0\;\\
0&1&0\\
0& 0&\;N\end{array}\right) ,\quad {\rm with}\quad N\in SO(n-1)\>,
\end{equation}
which means that $H^{(+)}=H^{(-)}$, and that both are isomorphic to $SO(n-1)$. Then, following the approach of~\cite{BPS} and Section~\ref{LAGSSSG}, the Lagrangian formulation of the resulting SSSG equations is provided by a gWZW action associated to the coset $SO(1,n-1)/SO(n-1)$ with a potential term. Again, in this case the reduction follows the pattern summarised by figure~\ref{Figura}.

The simplest case corresponds to $n=2$, where \begin{equation}
\Lambda_+=\Lambda_-=\widehat{k}[(1,0)]=\left(\begin{array}{ccc}
0 & -1&0\;\\
1&0&0\\
0& 0&0
\end{array}\right)
\end{equation}
and $H^{(+)}=H^{(-)}$ are trivial. Using the parameterisation~(\ref{AdS2}) for the field $\gamma$, the resulting equation is the sinh-Gordon equation
\begin{equation}
\partial_+\partial_-\chi +\mu_+\mu_-\sinh\chi=0.
\end{equation}

Again, the first case with non-trivial gauge groups $H^{(+)}$ and $H^{(-)}$ corresponds to $AdS_3$. Then
\begin{equation}
\Lambda_+=\Lambda_-=k[(1,0,0)]=\left(
\begin{array}{cccc}
0&-1&0&0\\
1&0&0&0\\
0&0&0&0\\
0&0&0&0
\end{array}\right)
\end{equation} 
and, using the notation~(\ref{BoostRot}),
the centraliser of $\Lambda_+=\Lambda_-$ in $SO(1,2)$ is now
\begin{equation}
H^{(+)}=H^{(-)}=\{{\rm e}^{\>x\> r}; \;x\in {\fl R} \} 
\simeq SO(2),
\end{equation}
which motivates the parameterisation
\begin{equation}
\gamma = {\rm e}^{\>\alpha\> r}\> {\rm e}^{\>\chi\> b_1}\> {\rm e}^{\>-\beta\> r} \in SO(1,2),
\end{equation}
where the three fields $\alpha$, $\beta$ and $\chi$ are real.
Correspondingly, the gauge fields in~(\ref{RedGauge}) are
\begin{equation}
A_-^{(R)} = a_-\> r\quad {\rm and}\quad A_+^{(L)} = a_+\> r,
\end{equation}
with $a_\pm \in{\fl R}$, and the $SO(2)_L\times SO(2)_R$ gauge transformations~(\ref{ESSSG1}) read
\begin{equation}
\alpha\rightarrow \alpha + \rho_-, \quad
\beta\rightarrow \beta + \rho_+, \quad{\rm and}\quad
a_\pm \rightarrow a_\pm -\partial_\pm \rho_{\mp},
\label{GaugeAdStime}
\end{equation}
where $h_\pm =  {\rm e}^{\>\rho_\pm r}$. In terms of the gauge invariant fields $\chi$, $b_+= a_+ +\partial_+\alpha$ and $b_-= a_- +\partial_-\beta$, the zero-curvature equations of motion~(\ref{ESSSG2}) are
\begin{eqnarray}
&&\partial_+\partial_-\chi - (b_+b_- - \mu_+\mu_-)\sinh\chi=0
\label{AdStimeone}\\[5pt]
&&\partial_+\bigl((1+\cosh\chi)b_-\bigr) -\partial_-\bigl((1+\cosh\chi)b_+\bigr)=0
\label{AdStimetwo}\\[5pt]
&&\partial_+\bigl((1-\cosh\chi)b_-\bigr) +\partial_-\bigl((1-\cosh\chi)b_+\bigr)=0,
\label{AdStimethree}
\end{eqnarray}
which are related to the SSSG equations~(\ref{Sone}--\ref{Sthree}) corresponding to $SO(3)$ by means of the analytic continuation $\theta\rightarrow i\chi$.
These equations admit a Lagrangian formulation in terms of a  $SO(1,2)/SO(2)$ gWZW action with a potential term that, again, gives rise to two local Lagrangians of the form~(\ref{AdSspaceLag}). They can be found by repeating the procedure used for~(\ref{AdSspaceone}--\ref{AdSspacethree}).

First, we use~(\ref{AdStimetwo}) to write
\begin{equation}
b_\pm = {2\over 1+\cosh\chi}\> \partial_\pm \phi,
\end{equation}
so that~(\ref{AdStimeone}) and~(\ref{AdStimethree}) become
\begin{eqnarray}
&&
\partial_\mu \bigl(\tanh^2(\chi/2) \partial^\mu\phi\bigr)=0\nonumber\\[5pt]
&&
\partial_+\partial_-\chi -{4\sinh\chi\over(1+\cosh\chi)^2}\> \partial_+\phi\partial_-\phi +\mu_+\mu_-\sinh\chi=0,
\label{CshGEoM}
\end{eqnarray}
which are the equations of motion of
\begin{equation} 
{\cal L}={1\over4} \partial_\mu\chi \partial^\mu\chi + \tanh^2(\chi/2)\partial_\mu\phi \partial^\mu\phi -\mu_+\mu_- \sinh^2(\chi/2).
\end{equation}
It corresponds to~(\ref{AdSspaceLag}) with 
\begin{equation}
U=i\sinh(\chi/2) {\rm e}^{\> i\phi}, \quad
V=i\sinh(\chi/2) {\rm e}^{\> -i\phi}
\quad {\rm and} \quad \lambda=+\mu_+\mu_-.
\label{AdsTimeLagD1}
\end{equation}
In a completely equivalent way, we can use~(\ref{AdStimethree}) to write
\begin{equation}
b_\pm =\pm {2\over 1-\cosh\chi}\> \partial_\pm \tilde\phi,
\end{equation}
which also leads to the equations of motion of~(\ref{AdSspaceLag}) but, now, 
\begin{equation}
U=\cosh(\chi/2) {\rm e}^{\> i\tilde\phi}, \quad
V=\cosh(\chi/2) {\rm e}^{\> -i\tilde\phi}
\quad {\rm and} \quad \lambda=-\mu_+\mu_-,
\label{AdsTimeLagD2}
\end{equation}
corresponding to
\begin{equation} 
{\cal L}={1\over4} \partial_\mu\chi \partial^\mu\chi +\coth^2(\chi/2)\partial_\mu\phi \partial^\mu\phi -\mu_+\mu_- \cosh^2(\chi/2).
\end{equation}
Remarkably, the two parameterisations~(\ref{AdsTimeLagD1}) and~(\ref{AdsTimeLagD2}) satisfy different reality conditions. The first one corresponds to
\begin{equation}
U=-V^\ast,
\end{equation}
and, as exhibited by~(\ref{CshGEoM}), the resulting Lagrangian is an generalization of the sinh-Gordon Lagrangian with an internal $U(1)$ degree of freedom; namely,
\begin{equation}
-{\cal L}\bigl[U,-U^\ast,\lambda\bigr]={\partial_\mu
U\>\partial^\mu U^\ast\over 1+ U U^\ast} + \lambda\>
U U^\ast\equiv
{\cal L}_{CShG},
\end{equation}
which is related to the complex sine-Gordon Lagrangian~(\ref{CSGLag})
by means of the analytic continuation
\begin{equation}
\psi \rightarrow U\quad {\rm and}\quad \psi^\ast \rightarrow -U^\ast
\end{equation}
or, equivalently, by $\theta\rightarrow i\chi$ in~(\ref{CSGLagD1}).
Correspondingly, the second parameterisation~(\ref{AdsTimeLagD2}) solves the reality conditions
\begin{equation}
U=+V^\ast \quad{\rm and}\quad |U|\geq1
\end{equation}
and, therefore, it describes solutions to the equations of motion of the complex sine-Gordon Lagrangian~(\ref{CSGLag}) with $|\psi|\geq1$. 

\subsection{`Lightlike' reduction: Bosonic strings on $AdS_n$}
\label{AdSLight}

Finally, we shall consider the SSSG equations specified by the constraints
\begin{equation}
T_{++} = T_{--} = 0,
\label{ChiralRed}
\end{equation}
whose solutions describe bosonic string configurations on $AdS_n$~\cite{DeVega,Jevicki}.
According to~(\ref{PCDads}), they correspond to
\begin{equation}
\Lambda_+=\Lambda_-=T^{(l)}.
\end{equation}
Obviously, the constraints~(\ref{ChiralRed}) do not break the classical conformal invariance of the original sigma model and, in fact, the SSSG equations resulting from this type of reduction do not follow the pattern summarised by Figure~\ref{Figura}.

In practice, most of the differences can be traced back to the fact that \begin{equation}
\mathop{\rm Ker} \bigl( {\mathop{\rm Ad}}_{T^{(l)}}\bigr) \cap \mathop{\rm Im} \bigl( {\mathop{\rm Ad}}_{T^{(l)}}\bigr) \not=\{0\},
\label{NonDisjoint}
\end{equation}
which is possible because $\mathop{\rm Tr}(T^{(l)\>2})=0$. 
Then, the zero-curvature condition~(\ref{SSSG3}) does not imply~(\ref{SSSG4}), which has two important direct consequences. The first one is that the gauge-fixing conditions~(\ref{SSSGLSgauge}) cannot be imposed and, therefore, the
resulting SSSG equations cannot be written as a system of NAAT equations. The second concerns the derivation of their Lagrangian formulation, which involves the gauge-fixing conditions~(\ref{EoMgWZW2}). In Appendix~\ref{AppGauge} we show that the consistency of those conditions relies on the identities~(\ref{SSSG4}). Therefore, the approach of~\cite{BPS} and Section~\ref{LAGSSSG}
cannot be used to derive a Lagrangian formulation for the SSSG equations corresponding to the constraints~(\ref{ChiralRed}).

Eq~(\ref{NonDisjoint}) can be easily verified by looking at the generic form of the elements in $\mathop{\rm Ker} \bigl( {\mathop{\rm Ad}}_{T^{(l)}}\bigr)$ and $\mathop{\rm Im} \bigl( {\mathop{\rm Ad}}_{T^{(l)}}\bigr)$, which is
\begin{equation}
\left(\begin{array}{cccc}
0\;&-p&0&p\;\\
p\;&0&\bfm{a}&0\\
0\;&\bfm{a}^{\rm T}&{\cal Q}&-\bfm{a}^{\rm T}\\
p\;&0&\bfm{a}&0
\end{array}\right)
\quad{\rm and}\quad
\left(\begin{array}{cccc}
0\;&-p&\bfm{s}&p\;\\
p\;&0&\bfm{a}&q\\
\bfm{s}^{\rm T}\;&\bfm{a}^{\rm T}&0&-\bfm{a}^{\rm T}\\
p\;&q&\bfm{a}&0
\end{array}\right),
\label{AdSKerIm}
\end{equation}
respectively, where~$p$ and~$q$ are real numbers, $\bfm{a}$ and~$\bfm{s}$ are real $(n-2)$-dimensional file vectors, and ${\cal Q}^{\rm T}=-{\cal Q}$ is a $(n-2)\times(n-2)$ skew-symmetric matrix. Clearly,
\begin{equation}
\left(\begin{array}{cccc}
0\;&-p&0&p\;\\
p\;&0&\bfm{a}&0\\
0\;&\bfm{a}^{\rm T}&0&-\bfm{a}^{\rm T}\\
p\;&0&\bfm{a}&0
\end{array}\right)\in \mathop{\rm Ker} \bigl( {\mathop{\rm Ad}}_{T^{(l)}}\bigr) \cap \mathop{\rm Im} \bigl( {\mathop{\rm Ad}}_{T^{(l)}}\bigr)\not=\{0\}
\end{equation}
for any value of $p$ and $\bfm{a}$. Furthermore,
\begin{equation}
\left(\begin{array}{cccc}
0\;&-r&0&-r\;\\
r\;&0&\bfm{a}&0\\
0\;&\bfm{a}^{\rm T}&0&\bfm{a}^{\rm T}\\
-r\;&0&-\bfm{a}&0
\end{array}\right)\notin \mathop{\rm Ker} \bigl( {\mathop{\rm Ad}}_{T^{(l)}}\bigr) \cup \mathop{\rm Im} \bigl( {\mathop{\rm Ad}}_{T^{(l)}}\bigr),
\end{equation}
which exhibits that the decompositions~(\ref{LambdaDec1}) are actually not satisfied in this case.

According to~(\ref{AdSKerIm}), the infinitesimal generators of the centraliser of $T^{(l)}=\widehat{k}[(1,0,\ldots,0,1)]$ in $SO(1,n-1)$ are of the form
\begin{equation}
\left(\begin{array}{cccc}
0\;&0&\cdots&0\;\\
0\;&0&\bfm{a}&0\\
\vdots\;&\bfm{a}^{\rm T}&{\cal Q}&-\bfm{a}^{\rm T}\\
0\;&0&\bfm{a}&0
\end{array}\right).
\label{GenCentLight}
\end{equation}
They generate the little group of the real $n$-dimensional vector ${\bf v}=(1,0,\ldots,0,1)^{\rm T}$ in $SO(1,n-1)$. Therefore, $H^{(+)}=H^{(-)}$, and they are isomorphic to the noncompact Euclidean group $E(n-2)$, which is the symmetry group of $(n-2)$-dimensional Euclidean space~\cite{Wigner}. Then, as explained in Section~\ref{SSSG}, the corresponding SSSG equations are zero curvature equations of the form~(\ref{ESSSG2}) defined on the left-right asymmetric coset $SO(1,n-1)/E_L(n-2)\times E_R(n-2)$.

We have already pointed out that the constraints~(\ref{ChiralRed}) do not break the conformal invariance of the original sigma model. In fact, the corresponding SSSG equations are invariant under the conformal transformation
\begin{eqnarray}
&x_\pm \rightarrow {\rm e}^{\>-\eta_\pm\>} x_\pm, \quad
\gamma \rightarrow {\rm e}^{\>-\eta_- B\>} \gamma {\>\rm e}^{\>\eta_+ B},&
\nonumber\\[5pt]
&A_+^{(L)}\rightarrow {\rm e}^{\>\eta_+\>} {\rm e}^{\>-\eta_- B\>} A_+^{(L)} {\rm e}^{\>\eta_- B\>}, \quad 
A_-^{(R)}\rightarrow {\rm e}^{\>\eta_-\>} {\rm e}^{\>-\eta_+ B\>} A_-^{(R)} {\rm e}^{\>\eta_+ B\>},&
\label{ConfTrans}
\end{eqnarray}
where $\eta_+=\eta_+(x_+)$ and $\eta_-=\eta_-(x_-)$ are real-valued chiral functions, and
\begin{equation}
B=\left(\begin{array}{cccc}
0 & 0&\cdots&0\\
0 & 0&\cdots&1\\
\vdots&\vdots&\vdots&\vdots\\
0 & 1&\cdots&0
\end{array}\right)\quad {\rm satisfies}\quad [B,T^{(l)}]=T^{(l)}.
\label{ConfGen}
\end{equation}
This can be checked as follows. Let us write the zero-curvature condition~(\ref{ESSSG2}) as $[{\cal L}_+,{\cal L}_-]=0$, where
\begin{equation}
{\cal L}_+= \partial_+ + \gamma^{-1}\partial_+\gamma + \gamma^{-1} A_+^{(L)} \gamma + z \mu_+\> \Lambda_+
\quad{\rm and}\quad
{\cal L}_-= \partial_- + A_-^{(R)} + z^{-1} \mu_- \>\gamma^{-1} \Lambda_-\gamma.
\label{Lax}
\end{equation}
Then, under~(\ref{ConfTrans}), ${\cal L}_+$ and ${\cal L}_-$ transform as
\begin{equation}
{\cal L}_+\rightarrow {\rm e}^{\>\eta_+\>} {\rm e}^{\>-\eta_+ B\>} {\cal L}_+ {\rm e}^{\>\eta_+ B\>}
\quad{\rm and}\quad
{\cal L}_-\rightarrow {\rm e}^{\>\eta_-\>} {\rm e}^{\>-\eta_+ B\>} {\cal L}_- {\rm e}^{\>\eta_+ B\>},
\end{equation}
which explicitly preserve the form of the zero-curvature condition.
It is worth noticing that, in terms of the reduced currents~(\ref{RedCurrents}) and taking~(\ref{GaugeJ}) into account, the conformal transformation~(\ref{ConfTrans}) corresponds simply to $J_\pm\rightarrow {\rm e}^{\>\eta_\pm\>} J_\pm$~\footnote{Since $\Lambda_+=\Lambda_-=T^{(l)}$, this can be easily checked as follows
\begin{eqnarray*}
&&
J_-=\mu_-\gamma^{-1}\Lambda_- \gamma 
\>\buildrel \mbox{(\ref{ConfTrans})} \over{\hbox to 30pt{\rightarrowfill}}\>
{\rm e}^{\>-\eta_+ B\>}({\rm e}^{\>\eta_-\>}J_-){\rm e}^{\>\eta_+ B\>}
\>\buildrel \mbox{(\ref{GaugeJ})} \over{\hbox to 30pt{\rightarrowfill}}\>
{\rm e}^{\>\eta_-\>}J_-,
\nonumber\\[5pt]
&&
{\rm and}\quad J_+=\mu_+\Lambda_+
\>\buildrel \mbox{(\ref{GaugeJ})} \over{\hbox to 30pt{\rightarrowfill}}\>
{\rm e}^{\>\eta_+ B\>}J_+{\>\rm e}^{\>-\eta_+ B\>}
={\rm e}^{\>\eta_+\>}J_+.
\end{eqnarray*}
}.

The simplest example corresponds to $n=2$, where $H^{(+)}$ and $H^{(-)}$ are trivial and the parameterisation of the field $\gamma$ is given by~(\ref{AdS2}). Then,
\begin{equation}
\Lambda_+=\Lambda_-=\widehat{k}[(1,1)]=\left(\begin{array}{ccc}
0 & -1&1\;\\
1&0&0\\
1& 0&0
\end{array}\right),
\end{equation}
and the resulting SSSG equation is the trivial one,  $\partial_+\partial_-\chi=0$. However, taking~(\ref{AdS2plus}) into account, for $AdS_2$ there is a second, non-equivalent solution to the constraints~(\ref{ChiralRed}); namely,
\begin{equation}
\Lambda_+=T^{(l)}=\widehat{k}[(1,1)]=\left(\begin{array}{ccc}
0 & -1&1\;\\
1&0&0\\
1& 0&0
\end{array}\right)
\quad{\rm and}\quad
\Lambda_-=\widetilde{T}^{(l)}=\widehat{k}[(1,-1)]=\left(\begin{array}{ccc}
0 & -1&-1\;\\
1&0&0\\
-1& 0&0
\end{array}\right).
\label{AdS2extra}
\end{equation}
As noticed originally in~\cite{GTseytlin2}, it leads to the SSSG equation
\begin{equation}
\partial_+\partial_-\chi + 2 \mu_+\mu_- {\rm e}^{\>\chi}=0,
\label{Liouville}
\end{equation}
which is the well known Liouville equation whose Lagrangian is
\begin{equation}
{\cal L}=\partial_+\chi \partial_-\chi -2 \mu_+\mu_- {\rm e}^{\>\chi}.
\end{equation}
Since $[B,\widetilde{T}^{(l)}] = -\widetilde{T}^{(l)}$, the transformation~(\ref{ConfTrans}) reads now
\begin{eqnarray}
&x_\pm \rightarrow {\rm e}^{\>-\eta_\pm\>} x_\pm, \quad
\gamma \rightarrow {\rm e}^{\>\eta_- B\>} \gamma {\>\rm e}^{\>\eta_+ B},&
\nonumber\\[5pt]
&A_+^{(L)}\rightarrow {\rm e}^{\>\eta_+\>} {\rm e}^{\>+\eta_- B\>} A_+^{(L)} {\rm e}^{\>-\eta_- B\>}, \quad 
A_-^{(R)}\rightarrow {\rm e}^{\>\eta_-\>} {\rm e}^{\>-\eta_+ B\>} A_-^{(R)} {\rm e}^{\>\eta_+ B\>}&
\label{ConfTransB}
\end{eqnarray}
and, in this case, it corresponds to $\chi\rightarrow \chi+\eta_+ +\eta_-$, which summarises the conformal symmetry of~(\ref{Liouville}).

The first case with non-trivial gauge groups $H^{(+)}$ and $H^{(-)}$ is the reduction of $AdS_3$ with $\Lambda_+=\Lambda_-=k[(1,0,1)]$. However, it will be more useful to consider the equivalent choice
\begin{equation}
\Lambda_+=k[(1,0,1)]=\left(
\begin{array}{cccc}
0&-1&0&1\\
1&0&0&0\\
0&0&0&0\\
1&0&0&0
\end{array}\right)
\quad{\rm and}\quad 
\Lambda_-=k[(1,0,-1)]=\left(
\begin{array}{cccc}
0&-1&0&-1\\
1&0&0&0\\
0&0&0&0\\
-1&0&0&0
\end{array}\right),
\label{ChoiceAdS3}
\end{equation}
whose centralisers in $SO(1,2)$ are
\begin{equation}
H^{(+)}=\{{\rm e}^{\>x\> g_+}; \;x\in {\fl R} \} 
\simeq E(1) 
\quad{\rm and}\quad
H^{(-)}=\{{\rm e}^{\>x\> g_-}; \;x\in {\fl R} \} 
\label{E1centraliser}
\simeq E(1),
\end{equation}
with
\begin{equation}
g_+=\left(
\begin{array}{cccc}
0&0&0&0\\
0&0&1&0\\
0&1&0&-1\\
0&0&1&0
\end{array}\right)
\quad{\rm and}\quad
g_-=\left(
\begin{array}{cccc}
0&0&0&0\\
0&0&1&0\\
0&1&0&1\\
0&0&-1&0
\end{array}\right).
\end{equation}
It is worth noticing that $g_\pm^3=0$ and, consequently,
\begin{equation}
{\rm e}^{\>x\> g_+} =\left(
\begin{array}{cccc}
1&0&0&0\\
0&1+{x^2\over2}&\>x\>&-{x^2\over2}\\
0&x&1&-x\\
0&{x^2\over2}&x&1-{x^2\over2}
\end{array}\right)
\quad{\rm and}\quad
{\rm e}^{\>x\> g_-} =\left(
\begin{array}{cccc}
1&0&0&0\\
0&1+{x^2\over2}&\>x\>&{x^2\over2}\\
0&x&1&x\\
0&-{x^2\over2}&\>-x\>&1-{x^2\over2}
\end{array}\right).
\end{equation}
Eq.~(\ref{E1centraliser}) motivates the parameterisation~\begin{equation}
\gamma = {\rm e}^{\>\alpha\> g_-}\> {\rm e}^{\>\chi\> B}\> {\rm e}^{\>-\beta\> g_+} \in SO(1,2),
\end{equation}
where the fields $\alpha$, $\beta$ and $\chi$ are real, and $B$ is given by~(\ref{ConfGen}).~\footnote{ 
It can be easily checked that $\Lambda_- = \gamma^{-1}\Lambda_+ \gamma$ for $\beta=\alpha$ and $\chi= - \ln \alpha^2$, which confirms that choosing $\Lambda_+=\Lambda_-=k[(1,0,1)]$ is indeed equivalent to~(\ref{ChoiceAdS3}). In particular, if $\alpha^2\rightarrow+\infty$, then $\gamma\rightarrow\mathop{\rm diag}(1,1,-1,-1)$, which is in (the identity component of) $SO(1,2)\subset SO(2,2)$.}
Correspondingly, the gauge fields in~(\ref{RedGauge}) are
\begin{equation}
A_-^{(R)} = a_-\> g_+\quad {\rm and}\quad A_+^{(L)} = a_+\> g_-,
\end{equation}
with $a_\pm \in{\fl R}$, and the $E(1)_L\times E(1)_R$ gauge transformations~(\ref{ESSSG1}) read
\begin{equation}
\alpha\rightarrow \alpha + \rho_-, \quad
\beta\rightarrow \beta + \rho_+, \quad{\rm and}\quad
a_\pm \rightarrow a_\pm -\partial_\pm \rho_{\mp},
\label{GaugeAdSlight}
\end{equation}
where $h_\pm =  {\rm e}^{\>\rho_\pm g_\pm}$.  In terms of the gauge invariant fields $\chi$, $b_+= a_+ +\partial_+\alpha$ and $b_-= a_- +\partial_-\beta$, the zero-curvature equations of motion~(\ref{ESSSG2}) are now
\begin{eqnarray}
&&
\partial_+\partial_-\chi + 2{\rm e}^{\>\chi} \bigl(b_+b_- + \mu_+\mu_-\bigr)=0
\label{AdSlightone}
\\[5pt]
&&
\partial_-({\rm e}^{\>\chi}b_+)=\partial_+({\rm e}^{\>\chi}b_-)=0.
\label{AdSlighttwo}
\end{eqnarray}
These equations exhibit the conformal symmetry summarised by~(\ref{ConfTransB}), which in this case reads
\begin{equation}
x_\pm \rightarrow {\rm e}^{\>-\eta_\pm\>} x_\pm, \quad
\chi \rightarrow \chi+\eta_+ +\eta_- \quad{\rm and}\quad
b_\pm \rightarrow {\rm e}^{\pm(\eta_+-\eta_-)} b_\pm.
\label{ConfTransC}
\end{equation}

The usual way to deal with~(\ref{AdSlightone}--\ref{AdSlighttwo}) is to explicitly break conformal invariance by considering a particular solution to the two equations~(\ref{AdSlighttwo})~\cite{DeVega,Jevicki},
\begin{equation}
b_+=\mu_+{\rm e}^{-\chi\>} u(x_+)
\quad{\rm and}\quad
b_-=\mu_-{\rm e}^{-\chi\>} v(x_-).
\end{equation}
Then,~(\ref{AdSlightone}) becomes
\begin{equation}
\partial_+\partial_-\chi + 2\mu_+\mu_-\left({\rm e}^{\>\chi} + u(x_+)v(x_-) {\rm e}^{-\chi}\right)=0,
\end{equation}
which can be transformed either in the sinh-Gordon equation \mbox{$\partial_+\partial_-\chi + 4\mu_+\mu_-\sinh\chi=0$} or the cosh-Gordon equation $\partial_+\partial_-\chi + 4\mu_+\mu_-\cosh\chi=0$ by means of the conformal transformation
\begin{equation}
x_\pm \rightarrow {\rm e}^{\>-\widetilde{\eta}_\pm\>} x_\pm
\quad{\rm and}\quad \chi \rightarrow \chi+\widetilde{\eta}_+ +\widetilde{\eta}_-
\label{ConfTransD}
\end{equation}
with ${\rm e}^{\>2\widetilde{\eta}_+}=|u(x_+)|$ and ${\rm e}^{\>2\widetilde{\eta}_-}=|v(x_-)|$.
Nevertheless, it would be more satisfactory to find a conformal invariant Lagrangian action whose equations of motion are~(\ref{AdSlightone}--\ref{AdSlighttwo}).
As explained at the beginning of this section, the approach of~\cite{BPS} and Section~\ref{LAGSSSG} cannot be applied to the `lightlike' reduction of the $AdS_n$ sigma model and, in fact, finding the Lagrangian formulation of this type of SSSG equations remains a open problem.

\section{Conclusions}
\label{Conclusions}

In this paper we have presented a systematic group theoretical formulation of the Pohlmeyer reduction of two-dimensional nonlinear sigma models with target-space a symmetric space $F/G$. The reduction consists in constraining all the chiral densities that display the classical conformal invariance of the sigma model to take constant values. This provides a map between the equations of motion of the sigma models and a class of integrable multi-component generalisations of the sine-Gordon equation known as symmetric space sine-Gordon (SSSG) equations.
Each set of SSSG equations is specified by a triplet of data $(F/G,\Lambda_+,\Lambda_-)$, where $\Lambda_+$ and $\Lambda_-$ are constant elements in a maximal abelian subspace, say~${\go a}$, of the 
orthogonal complement of the Lie algebra ${\go g}$ of $G$ in the Lie algebra ${\go f}$ of $F$. Then, $H^{(+)}$ and $H^{(-)}$ are the centralisers of $\Lambda_+$ and $\Lambda_-$ in $G$, respectively, and 
the equations are written as zero-curvature conditions on the left-right asymmetric coset space $G/H_L^{(-)}\times H_R^{(+)}$.
For particular gauge fixing conditions, they take the form of non-abelian affine Toda equations, which is how they usually appear in the literature~\cite{ECPn,EHCPn,SSSG2,SSSG3,SSSG4,BPS}.

The Lagrangian formulation of the SSSG equations was a long-standing problem until Bakas, Park and Shin proposed their identification with the equations of motion of specific gauged Wess-Zumino-Witten (gWZW) actions modified by suitable potentials~\cite{BPS} (see also~\cite{Parkold,HMP,FMSG,HSG}). 
This Lagrangian formulation suggests a perturbed conformal field theory approach to the quantization of these integrable systems.
Moreover, it is one of the key ingredients of a recent proposal to find a novel manifestly two-dimensional Lorentz invariant formulation of superstring theory on $AdS_5\times S^5$~\cite{GTseytlin1,GTseytlin2,MikhailovSS}.
The original construction of~\cite{BPS} was restricted to the cases with $H^{(+)}=H^{(-)}$, but we have shown that the SSSG equations admit a Lagrangian formulation in terms of a gauged WZW action with a potential term if both $H^{(+)}$ and $H^{(-)}$ are isomorphic to a single Lie group $H$. Remarkably, the equations with $H^{(+)}\not=H^{(-)}$ correspond to integrable perturbations of asymmetric coset models whose spectrum includes massive and massless modes.

As pointed out in~\cite{GTseytlin1}, the Lagrangian formulation
in terms of a gauge WZW action with a potential term involves also a particular choice of gauge fixing conditions, whose consistency has been clarified.
Our results also show that the Lagrangian action, which is associated to the coset $G/H$, is not unique. The different actions are related by $H_L^{(-)}\times H_R^{(+)}$ gauge transformations, but from the point of view of the Lagrangian actions themselves those relations take the form of non-trivial target-space duality transformations similar to those discussed in~\cite{HSGphases} whose structure should be clarified.
Our systematic formulation also makes explicit the relation between the degrees of freedom of the original nonlinear sigma model and those of the relevant Lagrangian actions. It is summarised by Figure~\ref{Figura}.

In general, a single symmetric space may give rise to different sets of SSSG equations. Their number depends both on the type of symmetric space and on its rank, which is the dimension of the abelian subspace ${\go a}$, where $\Lambda_+$ and $\Lambda_-$ live.
When the symmetric space is of definite signature and rank~1, the only independent constraints are \mbox{$T_{++} =\mu_+^2>0$} and $T_{--} =\mu_-^2>0$, and the reduction procedure gives rise to a single set of SSSG equations. These constraints can be identified with the Virasoro constraints of bosonic string theory on ${\fl R}_t\times {\cal M}$, and the solutions to the corresponding SSSG equations provide string configurations moving on curved space-times of this type. This has been widely used to construct string configurations on the ${\fl R}_t\times S^n$ subspaces of $AdS_5\times S^5$ in the context of the investigation of the AdS/CFT correspondence~\cite{HM,CDO}. Hopefully, it will also
help with the construction of string configurations on the subspaces of $AdS_4 \times {\fl C}P^3$ that could be relevant to investigate the recently proposed
duality between superstrings moving on this space-time and
$N = 6$ super Chern-Simons theory~\cite{NEW}.
In contrast, if $\mathop{\rm rank}(F/G)>1$ one has to constrain other chiral densities to be constant in addition to $T_{++}$ and $T_{++}$, and for different values of those constants the reduction procedure gives rise to rather different sets of SSSG equations. Their solutions should correspond to special bosonic string configurations that satisfy additional constraints whose interpretation would be interesting to investigate.
In both cases, it is worth mentioning that the reduction procedure does not fix the sign of $\mu_+$ and $\mu_-$, which leaves free the overall sign of the potential term in the Lagrangian formulation. This might be important since, in some cases, the SSSG equations have different soliton solutions for each sign. The simplest example is provided by the complex sine-Gordon equation that exhibits two different types of soliton solutions~\cite{Lund,Getmanov} (see also~\cite{HSGphases}).

The case of the symmetric spaces of indefinite signature is much more complicated, and we have made no attempt to discuss the corresponding SSSG equations in general. Instead,  we have only considered the reduction of sigma models with target space an anti-de~Sitter space $AdS_n$, which has Lorentzian signature and rank~1. 
In this case, the relevant constraints are $T_{++} =\lambda_+$ and $T_{--} =\lambda_-$ but, as a consequence of the indefinite signature of $AdS_n$, the sign of $\lambda_+$ and $\lambda_-$ is free and the resulting SSSG equations are different for each sign. We have distinguished three basic types of reductions. The first one, called here `spacelike', corresponds to $\lambda_+,\lambda_->0$ and gives rise to SSSG equations with a (non-unique) Lagrangian formulation in terms of a gWZW action corresponding to the coset $SO(1,n-1)/SO(1,n-2)$ with a potential term.
The solutions to these equations describe bosonic string configurations on ${\fl R}_t\times AdS_n$. The second, named `timelike', corresponds to \mbox{$\lambda_+,\lambda_-<0$}. It gives rise to SSSG equations which are the equations of motion of a (non-unique) gWZW action corresponding to $SO(1,n-1)/SO(n-1)$ with a potential term. 
Their solutions describe bosonic string configurations on $AdS_n\times S^1_\vartheta$. This second type of reduction is the relevant one in the new formulation of superstring theory on $AdS_5\times S^5$ proposed in~\cite{GTseytlin1,GTseytlin2,MikhailovSS}.
The details of these two types of reductions follow the pattern summarised by Figure~\ref{Figura}.
The third type, named `lightlike, is rather different to the others. It corresponds to $\lambda_+=\lambda_-=0$ which, clearly, does not break the classical conformal invariance of the sigma model. The corresponding SSSG equations take the form of zero-curvature conditions on the left-right asymmetric coset space $SO(1,n-1)/E(n-2)_L\times E(n-2)_R$, where $E(n-2)$ is the noncompact symmetry group of $(n-2)$-dimensional Euclidean space. In this case, neither the approach of~\cite{BPS} nor our generalization in Section~\ref{LAGSSSG} can be used to find a Lagrangian formulation, which remains an open problem. The solutions to these equations have already been used to construct bosonic string configurations on $AdS_n$ in~\cite{DeVega,Jevicki}.

\newpage

\acknowledgments
I would like to thank LPTHE~(Paris~6--CNRS--Paris~7) for the kind hospitality while this work was in progress.
I am grateful to Arkady Tseytlin for explaining his work and for many useful comments. I also thank
Olivier Babelon, Tim Hollowood, Francesco Ravanini and Alfonso V\'azquez Ramallo for helpful discussions. This work was partially  supported by MEC (Spain) 
and FEDER (grants FPA2005-00188 and 
FPA2005-01963), by a INFN-MEC cooperation grant, by Xunta de Galicia (Conseller\'\i a de 
Educaci\'on and grant PGIDIT06PXIB296182PR), and by the 
Spanish Consolider-Ingenio 2010
Programme CPAN (CSD2007-00042).

\vspace{0.5 cm}
\appendix
\section{Consistency of the gauge conditions~(3.41)}
\label{AppGauge}

We shall investigate the conditions required to ensure that
for each solution $(\gamma,A_+^{(L},A_-^{(R)})$ to the equations of motion~(\ref{ESSSG2}) there is a gauge transformation of the form~(\ref{ESSSG1}) such that the transformed solution satisfies the conditions~(\ref{EoMgWZW2}).

First of all, we will write the conditions~(\ref{EoMgWZW2}) in terms of quantities that take values in ${\go h}$. Namely, using~(\ref{FinalGauge}), 
$$
A_+^{(L)}= \epsilon_L({\cal A}_+)\quad {\rm and}\quad
A_-^{(R)}= \epsilon_R({\cal A}_-),
$$
and introducing the notation
\begin{eqnarray}
&&
{\mathop{\rm P}}_{{\go h}_-}\Bigl(-\partial_-\gamma \gamma^{-1} + \gamma \epsilon_R({\cal A}_-)\gamma^{-1}\Bigr) = \epsilon_L(\Gamma_-)\nonumber\\[5pt]
&&
{\rm and}\quad
{\mathop{\rm P}}_{{\go h}_+}\Bigl(\gamma^{-1}\partial_+\gamma  + \gamma^{-1} \epsilon_L({\cal A}_+)\gamma\Bigr) = \epsilon_R(\Gamma_+),
\end{eqnarray}
with $\Gamma_+,\Gamma_-\in{\go h}$, the conditions~(\ref{EoMgWZW2}) become simply
\begin{equation}
\Gamma_--{\cal A}_- =\Gamma_+-{\cal A}_+=0.
\label{GaugeApp}
\end{equation}
Now, writing $h_-=\epsilon_L(h)$ and $h_+=\epsilon_R(\tilde{h})$, with $h,\tilde{h}\in H$, the gauge transformations~(\ref{ESSSG1}) lead to
\begin{eqnarray}
&&
\Gamma_+ -{\cal A}_+\rightarrow \tilde{h}(\Gamma_+ + \partial_+) \tilde{h}^{-1} - h({\cal A}_+ + \partial_+) h^{-1}
\nonumber\\[5pt]
&&
{\rm and}\quad
\Gamma_- -{\cal A}_- \rightarrow h(\Gamma_- + \partial_-) h^{-1} - \tilde{h}({\cal A}_- + \partial_-) \tilde{h}^{-1}.
\label{App2}
\end{eqnarray}
Therefore, the interpretation of~(\ref{GaugeApp}) as gauge conditions requires that there exists $h$ and $\tilde{h}$ such that the right-hand-sides of the equations~(\ref{App2}) vanish. This is equivalent to
the set of linear equations
\begin{eqnarray}
&&
\partial_+(\tilde{h}^{-1} h) = -\Gamma_+ (\tilde{h}^{-1} h) + (\tilde{h}^{-1} h) {\cal A}_+
\nonumber\\[5pt]
&&
{\rm and}\quad
\partial_-(\tilde{h}^{-1} h) = (\tilde{h}^{-1} h)\Gamma_- -  {\cal A}_- (\tilde{h}^{-1} h),
\label{App3}
\end{eqnarray}
whose  integrability conditions are
\begin{equation}
(\tilde{h}^{-1} h)\> [\partial_+ + {\cal A}_+, \partial_- + \Gamma_-] = [\partial_+ + \Gamma_+, \partial_- +  {\cal A}_-]\> (\tilde{h}^{-1} h)\>.
\end{equation}
They are trivially satisfied making use of~(\ref{SSSG4}), which ensures that
\begin{equation}
 [\partial_+ + \Gamma_+, \partial_- +  {\cal A}_-]=[\partial_+ + {\cal A}_+, \partial_- + \Gamma_-] =0\>.
\end{equation}
However,~(\ref{SSSG4}) holds provided that the decompositions~(\ref{LambdaDec1}) are valid. This is always true if the symmetric space is of definite signature, but it not always so in more general cases like the anti-de~Sitter spaces $AdS_n$ (see Section~\ref{AdSLight}).

\section{Pohlmeyer reduction of the $\bfm{S^3}$ sigma model}
\label{AppCSG}

Here we shall summarise the main features of the Pohlmeyer reduction of the $S^3$ nonlinear sigma model following the group theoretical approach of Section~\ref{SSSG}. The reduction of the $S^2$ and $S^3$ nonlinear sigma models was originally discussed by Pohlmeyer in~\cite{Pohl} using embedding coordinates. They lead to the equations of motion of the sine-Gordon and complex sine-Gordon models, respectively, which provide the pattern for all the other SSSG equations. In particular we shall emphasise the non-uniqueness of the Lagrangian formulation that, in this case, amounts simply to the freedom of choosing the sign of the potential term. The reduction of the ${\fl C}P^2$ sigma model discussed in Section~\ref{ExampleCPn} provides an example where the relationship between the two relevant Lagrangians is not so simple.

We start by setting our notation in general for the $n$-sphere $S^n=SO(n+1)/SO(n)$, which is a compact symmetric space of (definite signature and) type~I. Using the fundamental $(n+1)\times (n+1)$ representation of $SO(n+1)$ and its diagonally embedded $SO(n)$ subgroup, the elements $r\in{\go g}$ and $k\in{\go p}$ in~(\ref{SSalgebra})
are of the form
\begin{equation}
r=\left(
\begin{array}{cc}0 & 0 \\0 & {\cal N}\end{array}\right)
\quad {\rm and}\quad
k=\left(
\begin{array}{cc}
0&-\bfm{v}^{\rm T}\\
\bfm{v}&0
\end{array}\right)
\equiv k[\bfm{v}]
\end{equation}
where ${\cal N}=-{\cal N}^{\rm T}$ denotes a $n\times n$ skew-symmetric matrix and $\bfm{v}=(v_1,\ldots,v_n)^{\rm T}$ is a real $n$-dimensional column vector. It is easy to check that this decomposition is orthogonal with respect to the trace form and, moreover, that the rank of $S^n$ is~1. 

In order to motivate the generalization proposed in Section~\ref{AdS} for $AdS_n$, we shall rephrase the proof of the polar coordinate decomposition for $S^n$. Consider a generic element of $SO(n)\subset SO(n+1)$,
\begin{equation}
g=\left(
\begin{array}{cc}1 & 0 \\0 & N^{-1}\end{array}\right)
\quad{\rm with}\quad
N\in SO(n).
\end{equation}
Then, the transformation $k[\bfm{v}]\rightarrow g^{-1} k[\bfm{v}] g$ amounts to
$\bfm{v}\rightarrow N\bfm{v}$, 
which is just a $SO(n)$ rotation that preserves the quadratic form $-\mathop{\rm Tr}(k^2[\bfm{v}])/2=v_1^2+\cdots+v_n^2=|\bfm{v}|^2$. Therefore, for any fixed unitary vector $\bfm{e}_0\in {\fl R}^n$, it is well known that there is a matrix $N\in SO(n)$ such that $\bfm{v} N=|\bfm{v}|\>\bfm{e}_0$ or, equivalently, that there is $g\in SO(n)\subset SO(n+1)$ such that
\begin{equation}
g^{-1} k g = {1\over2}\sqrt{-\mathop{\rm Tr}(k^2)}\>
\left(\begin{array}{cc}0 & -\bfm{e}_0^{\rm T} \\\bfm{e}_0 & 0
\end{array}\right).
\end{equation}
This is just the polar coordinate decomposition used in Section~\ref{SSSG}. A convenient choice for the arbitrary unitary vector is $\bfm{e}_0^{\rm T}=(1,0,\ldots,0)$. Then, the polar coordinate decomposition for $S^n$ ensures that for each $k\in {\go p}$ there exists $\overline{g}\in SO(n)$ and $\mu\in {\fl R}$ such that
\begin{equation}
\overline{g}^{-1} k \overline{g} =\mu k[(1,0,\ldots,0)],
\label{PCDsimple}
\end{equation}
which should be compared with the generalised decomposition~(\ref{PCDads}) derived in Section~\ref{AdS} for $AdS_n$. It is worth noticing that, without loss of generality, $\mu$ can always be constrained to be positive in~(\ref{PCDsimple}), which is not always true in~(\ref{PCDads}). 

Then, in order to construct the SSSG equations corresponding to $S^3$, we choose
\begin{equation}
\Lambda_+=\Lambda_-=k[(1,0,0)]=\left(
\begin{array}{cccc}
0&-1& 0 & 0 \\
1&0&0&0\\
0&0&0&0\\
0&0&0&0
\end{array}\right)
\end{equation} 
in~(\ref{RedCurrents}). Moreover,
we will introduce the notation
\begin{equation}
r_1=\left(
\begin{array}{cccc}
0&0& 0 & 0 \\
0&0&0&0\\
0&0&0&1\\
0&0&-1&0
\end{array}\right)=-r_1^{\rm T}
\quad {\rm and}\quad
r_3=\left(
\begin{array}{cccc}
0&0& 0 & 0 \\
0&0&1&0\\
0&-1&0&0\\
0&0&0&0
\end{array}\right)=-r_3^{\rm T\>}
\end{equation}
so that the centraliser of $\Lambda_+=\Lambda_-$ in $SO(3)$ is
the one-parameter (abelian) group
\begin{equation}
H^{(+)}=H^{(-)}=\{{\rm e}^{\>x\> r_1}; \;x\in {\fl R} \} 
\simeq SO(2).
\label{SO2form}
\end{equation}
This motivates the use of the following parameterisation of Euler-angle type for the field $\gamma \in SO(3)\subset SO(4)$:
\begin{equation}
\gamma = {\rm e}^{\>\alpha\> r_1}\> {\rm e}^{\>\theta\> r_3}\> {\rm e}^{\>-\beta\> r_1}
\end{equation}
in terms of three real fields $\alpha$, $\beta$ and $\theta$. Correspondingly, the gauge fields in~(\ref{RedGauge}) are
\begin{equation}
A_-^{(R)} = a_-\> r_1\quad {\rm and}\quad A_+^{(L)} = a_+\> r_1,
\end{equation}
with $a_\pm \in{\fl R}$, and the $SO(2)_L\times SO(2)_R$ gauge transformations~(\ref{ESSSG1}) read
\begin{equation}
\alpha\rightarrow \alpha + \rho_-, \quad
\beta\rightarrow \beta + \rho_+, \quad{\rm and}\quad
a_\pm \rightarrow a_\pm -\partial_\pm \rho_{\mp},
\label{GaugeS3}
\end{equation}
where $h_\pm =  {\rm e}^{\>\rho_\pm r_1}$. Then, in terms of the gauge invariant fields
\begin{equation}
\theta,\quad b_+= a_+ +\partial_+\alpha\quad 
{\rm and}\quad b_-= a_- +\partial_-\beta, 
\end{equation}
the zero-curvature equations of motion~(\ref{ESSSG2}) are
\begin{eqnarray}
&&\partial_+\partial_-\theta - (b_+b_- - \mu_+\mu_-)\sin\theta=0
\label{Sone}\\[5pt]
&&\partial_+\bigl((1+\cos\theta)b_-\bigr) -\partial_-\bigl((1+\cos\theta)b_+\bigr)=0
\label{Stwo}\\[5pt]
&&\partial_+\bigl((1-\cos\theta)b_-\bigr) +\partial_-\bigl((1-\cos\theta)b_+\bigr)=0
\label{Sthree}.
\end{eqnarray}
The relationship with the equations of motion of the complex sine-Gordon Lagrangian
\begin{equation}
 {\cal L}_{\rm CSG}[\psi,\lambda]={\partial_\mu
\psi\>\partial^\mu \psi^\ast\over 1- \psi \psi^\ast} - \lambda\>
\psi \psi^\ast
\label{CSGLag}
\end{equation}
can be obtained in two different ways. First, we can use~(\ref{Stwo}) to write $b_+$ and $b_-$ in terms of a new field $\phi$; namely,
\begin{equation}
b_\pm = {2\over 1+\cos\theta}\> \partial_\pm \phi.
\label{Integ1}
\end{equation}
Then,~(\ref{Sone}) and~(\ref{Sthree}) become
\begin{eqnarray}
&&
\partial_\mu \bigl(\tan^2(\theta/2) \partial^\mu\phi\bigr)=0\nonumber\\[5pt]
&&
\partial_+\partial_-\theta -{4\sin\theta\over(1+\cos\theta)^2}\> \partial_+\phi\partial_-\phi +\mu_+\mu_-\sin\theta=0,
\label{Otra}
\end{eqnarray}
which are the equations of motion of
\begin{eqnarray} 
{\cal L}&=&{1\over4} \partial_\mu\theta \partial^\mu\theta + \tan^2(\theta/2)\partial_\mu\phi \partial^\mu\phi - \mu_+\mu_- \sin^2(\theta/2)\nonumber\\[5pt]
&=&
{\cal L}_{\rm CSG}[\sin(\theta/2) {\rm e}^{\> i\phi},+\mu_+\mu_-]\>.
\label{CSGLagD1}
\end{eqnarray} 
In a completely equivalent way, we can use~(\ref{Sthree}) to write
\begin{equation}
b_\pm =\pm {2\over 1-\cos\theta}\> \partial_\pm \tilde\phi,
\label{Integ2}
\end{equation}
which also leads to the equations of motion of~(\ref{CSGLag}) but, in this second case, 
\begin{eqnarray}
{\cal L}&=&
{1\over4} \partial_\mu\theta \partial^\mu\theta + \cot^2(\theta/2)\partial_\mu\tilde{\phi} \partial^\mu\tilde{\phi} +  \mu_+\mu_- \cos^2(\theta/2)\nonumber\\[5pt]
&=&
{\cal L}_{\rm CSG}[\cos(\theta/2) {\rm e}^{\> i\tilde{\phi}},-\mu_+\mu_-]\>.
\label{CSGLagD2}
\end{eqnarray}
Remarkably, all the solutions to the equation of motion of the complex sine-Gordon Lagrangian~(\ref{CSGLag}) that correspond to solutions to the SSSG equations associated to $S^3$ satisfy $|\psi|\leq1$. The solutions with $|\psi|\geq1$ provide solutions to the `timelike' SSSG equations associated to $AdS_3$ (see Section~\ref{Timelike}).

The two complex sine-Gordon Lagrangians~(\ref{CSGLagD1}) and~(\ref{CSGLagD2}) are related simply by means of
\begin{equation}
(\theta,\phi,+\mu_+\mu_-)\rightarrow (\pi-\theta,\tilde\phi,-\mu_+\mu_-)
\end{equation}
where, since $b_\pm$ and $\theta$ are the same in~(\ref{Integ1}) and in~(\ref{Integ2}),  
$\partial_\pm\tilde\phi = \pm \tan^2(\theta/2) \partial_\pm\phi$.
Remarkably, this is precisely the already known (on-shell) target-space duality transformation of the complex sine-Gordon Lagrangian~\cite{HSGphases,ParkDual} which, in this case, arises as a consequence of the fact that the solutions to the equations of motion of~(\ref{CSGLagD1}) and~(\ref{CSGLagD2}) describe the same system of SSSG equations.

As explained in Section~\ref{LAGSSSG}, the equations~(\ref{Sone}--\ref{Sthree}) admit a Lagrangian formulation in terms of a $SO(3)/SO(2)$ gWZW action with a potential term. It requires to reduce the $SO(2)_L\times SO(2)_R$ gauge symmetry~(\ref{GaugeS3}) using the gauge conditions~(\ref{EoMgWZW2}), which depend on the choice of two homomorphisms $\epsilon_{L/R}:SO(2)\rightarrow SO(3)$ constrained by~(\ref{AnomalyFree}).
In this case, there are only two non-equivalent choices.
Using the same notation for the corresponding homomorphisms between the Lie algebras of $SO(2)$ and $SO(3)$, the first one is $\epsilon_L=\epsilon_R=1$. It leads to the constraints
\begin{equation}
(a_+ +\partial_+\alpha)\cos\theta -\partial_+\beta =a_+
\quad{\rm and}\quad
(a_- +\partial_-\beta)\cos\theta -\partial_-\alpha  =a_-,
\end{equation}
which can be solved as
\begin{equation}
b_\pm = \pm {1\over 1-\cos\theta}\> \partial_\pm(\alpha-\beta).
\label{NInteg2}
\end{equation}
Therefore, since $b_+$ and $b_-$ are $SO(2)_L\times SO(2)_R$ gauge invariant, the residual $SO(2)$ gauge transformations correspond to $\rho_+=\rho_-$ in~(\ref{GaugeS3}), and the relevant gWZW action in~(\ref{SSSGAction}) is the one constructed using gauge transformations of `vector type'. 
The second possible choice is $\epsilon_L=-\epsilon_R=1$. It leads to
\begin{equation}
(a_+ +\partial_+\alpha)\cos\theta -\partial_+\beta =- a_+
\quad{\rm and}\quad
(a_- +\partial_-\beta)\cos\theta -\partial_-\alpha  =-a_-,
\end{equation}
which can be solved as
\begin{equation}
b_\pm = {1\over 1+\cos\theta}\> \partial_\pm(\alpha+\beta).
\label{NInteg1}
\end{equation}
Then, the residual $SO(2)$ gauge transformations correspond to $\rho_+=-\rho_-$ in~(\ref{GaugeS3}), and the gWZW action in~(\ref{SSSGAction}) is constructed using gauge transformations of `axial type'.
The correspondence with the Lagrangian formulation in terms of the complex sine-Gordon Lagrangian~(\ref{CSGLag}) is provided by the comparison of~(\ref{NInteg2}) and~(\ref{NInteg1}) with~(\ref{Integ2}) and~(\ref{Integ1}), respectively. It is in agreement with the results of~\cite{HSGphases} where it is also shown that the two  Lagrangian formulations are related (off-shell) by a target-space duality transformation generated by the global $SO(2)$ symmetry of the Lagrangian action~(\ref{SSSGAction}).


\end{document}